\documentclass[12pt]{article}

\font\grande=cmr9.5 scaled \magstep4
\font\medio=cmr9.5 scaled \magstep2
\outer\def\beginsection#1\par{\medbreak\bigskip
      \message{#1}\leftline{\bf#1}\nobreak\medskip
\vskip-\parskip
      \noindent}
\usepackage{graphicx}

\begin{document}
\bibliographystyle {unsrt}

\titlepage

\begin{flushright}
\end{flushright}
\vspace{5mm}
\begin{center}
{\grande The scalar modes of the relic gravitons}\\
\vspace{1.5cm}
 Massimo Giovannini
 \footnote{Electronic address: massimo.giovannini@cern.ch}\\
\vspace{1cm}
{{\sl Department of Physics, 
Theory Division, CERN, 1211 Geneva 23, Switzerland }}\\
\vspace{0.5cm}
{{\sl INFN, Section of Milan-Bicocca, 20126 Milan, Italy}}
\vspace*{0.5cm}
\end{center}

\vskip 0.5cm
\centerline{\medio  Abstract}
In conformally flat background geometries  
the long wavelength gravitons can be described in the fluid approximation and they  induce scalar fluctuations both during inflation and in the subsequent radiation-dominated epoch.  While this effect is minute and suppressed for a de Sitter stage of expansion, the fluctuations of the energy-momentum pseudo-tensor of the graviton fluid lead to curvature perturbations that increase with time all along the post-inflationary evolution.  An explicit calculation of these effects is presented  for a standard thermal history and it is shown that the growth of the  curvature perturbations caused by the long wavelength modes is approximately compensated by the slope of the power spectra of the energy density, pressure and anisotropic stress of the relic gravitons.
\vskip 0.5cm

\noindent

\vspace{5mm}

\vfill
\newpage
\renewcommand{\theequation}{1.\arabic{equation}}
\setcounter{equation}{0}
\section{Introduction}
\label{sec1}
The  tensor modes of the geometry
 can be efficiently amplified in conformally flat space-times thanks the pumping 
action of the (extrinsic) curvature \cite{gris1}.  As a consequence,
relic gravitons are often regarded as a direct signature of any 
scenario positing an early variation of the background gravitational field:
typical examples are the conventional inflationary paradigm \cite{star,FP,inflsp}
and various completions of the concordance lore \cite{compl1,compl2}.

The energy and momentum of the gravitational field itself 
cannot be localized. This perspective is vividly described, for instance, 
in \cite{MTW} (see also \cite{weinberg}) and it rests on the validity of the 
equivalence principle. Since there is no unique expression for the energy-momentum tensor of the relic 
gravitons, we must instead deal with a 
variety of pseudo-tensors whose definitions are mathematically different but physically equivalent. 
The classic Landau and Lifshitz approach \cite{LL} stipulates that a valid energy-momentum pseudo-tensor 
can be obtained (in Minkowski space) from the second-order fluctuations of the Einstein tensor (see also \cite{isaacson1,abramo1}).
The same strategy pioneered in a Minkowski background can be appropriately 
extended to conformally flat space-times \cite{mg}. In a complementary perspective Ford and Parker \cite{ford1} suggested that 
the second-order action for the tensor modes of the geometry in a conformally flat space-time 
of Friedmann-Robertson-Walker type coincides with the action of two 
minimally coupled scalar fields (one for each tensor polarization): it is then plausible 
to argue that the energy-momentum tensor of each polarization of the graviton must coincide with 
the one of a minimally coupled scalar field. 

In the present investigation it is observed that the  
the fluctuations of the energy density, of the pressure and of the anisotropic stress of the relic gravitons 
produce  secondary scalar modes of the geometry. During inflation 
their contribution can be safely ignored: while the energy density of the inflaton is approximately constant the energy density 
of the gravitons is suppressed. 
Conversely, during the post-inflationary phase the contribution of the long wavelength gravitons to the curvature perturbations 
 increases both during radiation and during the matter-dominated epoch.  The  rationale of this effect is, in short, 
the following. The fluctuations of the energy-momentum pseudo-tensor decrease more slowly than the dominant component 
of the energy density, at least in the case of the conventional thermal history dominated by radiation and, later on, by dust.  
To gauge the robustness of the physical results, the analysis will be conducted within the two complementary parametrizations of the energy-momentum pseudo-tensor mentioned in the previous paragraph.

The relic graviton fluid and its fluctuations are introduced  in section \ref{sec2}.
In section \ref{sec3} we examine the evolution of the scalar modes induced by an effective fluid 
of relic gravitons. Section 
\ref{sec4}  contains the evaluation of the various power spectra and the explicit estimates of the curvature perturbations. The concluding remarks are 
collected in section \ref{sec5}. To avoid digressions, some 
technical results of the analysis have been collected in the appendix.

\renewcommand{\theequation}{2.\arabic{equation}}
\setcounter{equation}{0}
\section{The fluid of relic gravitons}
\label{sec2}
\subsection{Generalities}
While the energy and momentum of the gravitational field itself cannot be localized,  there exist consistent frameworks  
for the analysis of a  gravitational energy-momentum pseudo-tensor on a given background geometry 
(be it flat space-time or any other space-time). In the present situation we shall be 
chiefly concerned with conformally flat geometries\footnote{The metric tensor of the background metric is given, in the present context, by 
$\overline{g}_{\mu\nu} = a^2(\tau) \eta_{\mu\nu}$; $a(\tau)$ is the scale factor, $\tau$ denotes 
the conformal time coordinate and $\eta_{\mu\nu}$ is the Minkowski metric with 
signature $(+,\,-,\,-,\,-)$.}. Let us first consider the case where the energy-momentum of the sources (denoted hereunder by $T_{\mu\nu}$) vanishes; the Einstein equations can then be written in the form\footnote{The Planck length will be defined as 
 $\ell_{\rm P} = \sqrt{8\pi G}$ in units $\hbar= c =1$. } 
\begin{equation}
\delta^{(1)} {\mathcal G}_{\mu}^{\nu} = \ell_{\mathrm{P}}^2  \,{\mathcal U}_{\mu}^{\nu} ,
\label{EIN1}
\end{equation}
where ${\mathcal G}_{\mu}^{\nu} = R_{\mu}^{\nu} - \delta_{\mu}^{\nu} R/2$  is the Einstein tensor
written in terms of the Ricci tensor and of the Ricci scalar denoted, respectively, by $R_{\mu}^{\nu}$ and $R$.  
In Eq. (\ref{EIN1})  $\delta^{(1)} {\mathcal G}_{\mu}^{\nu}$ stands as a symbol for the first-order fluctuation of the Einstein tensor. 
By definition ${\mathcal U}_{\mu}^{\nu}$ is the energy-momentum tensor of the 
gravitational field itself and it depends on the first-order fluctuations of the Einstein tensor:
\begin{equation}
{\mathcal U}_{\mu}^{\nu} = \frac{1}{\ell_{\mathrm{P}}^2} \biggl[ \delta^{(1)} {\mathcal G}_{\mu}^{\nu} - {\mathcal G}_{\mu}^{\nu} \biggr].
\label{EIN2}
\end{equation}
The explicit form of ${\mathcal U}_{\mu}^{\nu}$ depends on the transformation properties 
of the fluctuations with respect to three-dimensional rotations and also on the specific background geometry. In general terms
the first-order fluctuations of the Einstein tensor can be written as
\begin{equation}
\delta^{(1)}{\mathcal G}_{\mu}^{\nu} = \delta_{\mathrm{t}}^{(1)}{\mathcal G}_{\mu}^{\nu}  + \delta_{\mathrm{v}}^{(1)}{\mathcal G}_{\mu}^{\nu}  
+ \delta_{\mathrm{s}}^{(1)}{\mathcal G}_{\mu}^{\nu},
\label{EIN3}
\end{equation}
where the subscripts at the right-hand side denote, respectively the tensor, vector and scalar fluctuations of ${\mathcal G}_{\mu}^{\nu}$.

For a conventional post-inflationary thermal history the vector modes 
can be neglected since they are always suppressed both during and after inflation\footnote{In the present investigation the attention shall be focussed on the concordance 
lore with standard post-inflationary thermal history. }. Thus, using Eq. (\ref{EIN3}), Eqs. (\ref{EIN1}) and (\ref{EIN2}) can be written as\footnote{Note that $T_{\mu}^{\nu}$ 
is the energy-momentum tensor of the matter field that has now been included for a consistent treatment of the scalar modes of the geometry.}:
\begin{eqnarray}
\delta_{\mathrm{t}}^{(1)}{\mathcal G}_{\mu}^{\nu} &=& 0,
\label{EIN4}\\
\delta_{\mathrm{s}}^{(1)}{\mathcal G}_{\mu}^{\nu} &=& \ell_{\mathrm{P}}^2 \biggl[  \delta_{\mathrm{s}}^{(1)}T_{\mu}^{\nu} + {\mathcal U}_{\mu}^{\nu} \biggr],
\label{EIN5}\\
{\mathcal U}_{\mu}^{\nu} &=& - \frac{1}{\ell_{\mathrm{P}}^2}  \delta_{\mathrm{t}}^{(2)}{\mathcal G}_{\mu}^{\nu},
\label{EIN6}
\end{eqnarray}
where the fluctuations of $T_{\mu}^{\nu}$ have been included by assuming, as in the conventional case, that $\delta_{\mathrm{t}}^{(1)}T_{\mu}^{\nu}=\delta_{\mathrm{v}}^{(1)}T_{\mu}^{\nu}=0$. The superscript at the right-hand side of Eq. (\ref{EIN6}) denotes the second-order 
fluctuation of the corresponding quantity while the subscript refers to the tensor nature of the fluctuations. 
The components of the energy-momentum pseudo-tensor ${\mathcal U}_{\mu}^{\nu}$ are not 
covariantly conserved. However,  since the Bianchi identity $\nabla_{\mu} {\cal G}_{\nu}^{\mu}=0$
 should be valid to all orders, we will also have to demand  $\delta_{\rm t}^{(2)} ( \nabla_{\mu} {\cal G}^{\mu}_{\nu}) =0$. 
The perturbation of the Bianchi identity implies that the relic graviton fluid, in the Landau-Lifshitz parametrization, 
possesses an effective bulk viscosity (see appendix \ref{appA} for a discussion 
of this point). Equation (\ref{EIN6}) defines the Landau-Lisfshitz pseudo-tensor in its general form.
Moreover, Eqs. (\ref{EIN4}), (\ref{EIN5})--(\ref{EIN6}) concisely summarize all the logical steps of the 
present discussion. In particular, the explicit form of Eq. (\ref{EIN4}) gives the standard evolution of the 
tensor fluctuations of the geometry while Eq. (\ref{EIN5}) formally accounts for the evolution of the scalar modes 
modified by the relic gravitons whose pseudo-tensor is given by Eq. (\ref{EIN6}). 

A complementary strategy for a suitable definition of the energy-momentum pseudo-tensor of the relic gravitons has been suggested in \cite{ford1}.
In a  conformally flat background $\overline{g}_{\mu\nu} = a^2(\tau) \eta_{\mu\nu}$, the Einstein-Hilbert action perturbed to second 
order in the amplitude of the tensor modes, 
reads, up to total derivatives, 
\begin{equation}
S_{\mathrm{gw}} = \delta_{\rm t}^{(2)} S = \frac{1}{8 \ell_{\mathrm{P}}} \int d^{4} x\,\,\sqrt{-\overline{g}} \,\,\overline{g}^{\alpha\beta} 
\partial_{\alpha} h_{ij}\partial_{\beta} h_{ij},
\label{FPdef1}
\end{equation}
where $ \overline{g}_{\mu\nu}(\tau) = a^{2}(\tau)\eta_{\mu\nu}$  and the first-order  fluctuation of the metric is defined as 
$g_{\mu\nu}(\vec{x},\tau) \to \overline{g}_{\mu\nu}(\tau)+ \delta^{(1)}_{\mathrm{t}}\,g_{\mu\nu}(\vec{x},\tau)$. Furthermore 
$ \delta^{(1)}_{\mathrm{t}}\,g_{0i}(\vec{x},\tau) =  \delta^{(1)}_{\mathrm{t}}\,g_{00}(\vec{x},\tau)=0$ and 
\begin{equation}
 \delta^{(1)}_{\mathrm{t}}\,g_{ij}(\vec{x},\tau) = - a^{2}(\tau)\, h_{ij}(\vec{x},\tau), \qquad \partial_{i} h^{i}_{j} = h_{i}^{i} =0.
\label{tens2}
\end{equation} 
The energy-momentum pseudo-tensor is formally obtained from Eq. (\ref{FPdef1}) by functional derivation 
with respect to $\overline{g}_{\mu\nu}$, i.e. 
\begin{equation}
\delta S= \frac{1}{2} \int d^{4} x\, \, \sqrt{-\overline{g}} \, {\mathcal W}_{\mu\nu}  \,\delta \overline{g}_{\mu\nu}.
\label{FPdef2}
\end{equation} 
As expected, by setting to zero the variation  of Eq. (\ref{FPdef1}) with respect to $h_{ij}$, we obtain the evolution 
of the amplitude of the tensor fluctuations in our conformally flat background geometry:
\begin{equation}
h_{ij}^{\prime\prime} + 2 {\mathcal H}\, h_{ij}^{\prime} - \nabla^2 h_{ij} =0,
\label{MFUN}
\end{equation}
where the prime denotes a derivation with respect to the conformal time coordinate $\tau$ and, as usual,  ${\mathcal H}= a^{\prime}/a$.
Equation (\ref{MFUN}) has the same content of Eq. (\ref{EIN4}) since $\delta_{\mathrm{t}}^{(1)} R =0$ and $\delta_{\mathrm{t}}^{(1)} {\mathcal G}_{0}^{0} = 
\delta_{\mathrm{t}}^{(1)} {\mathcal G}_{i}^{0}=0$.

\subsection{Explicit forms of the pseudo-tensors}

According to Eq. (\ref{EIN2}) the energy-momentum pseudo-tensor can be computed from the second-order fluctuations of the 
Einstein tensor after a straightforward but algebraically lengthy procedure. Some of the relevant results of this calculation
 are reported in the first paper of Ref. \cite{mg} and subsequently rederived
(see second paper of Ref. \cite{mg} and references therein). The main result for the components 
 of the energy-momentum pseudo-tensor can be expressed, in the conformal 
 time parametrization, as follows:
\begin{eqnarray}
&&{\mathcal U}_{0}^{0} =  \frac{1}{a^2 \ell_{\rm P}^2} \biggl[ {\cal H} \,
H_{k\ell }\, h^{k\ell} + \frac{1}{8} ( \partial_{m} h_{k\ell} \partial^{m} h^{k\ell} + 
H_{k\ell} H^{k\ell})\biggr],
\label{LL1}\\
&& {\mathcal U}_{i}^{j} = \frac{{\mathcal U}}{3} \delta_{i}^{j} + \Pi_{i}^{j}, \qquad  {\mathcal U}_{i}^{0} = \frac{1}{4 \ell_{\mathrm{P}}^2 a^2} H_{k\ell} \,\partial_{i} h_{k\ell}, 
\label{LL2}
\end{eqnarray}
where 
\begin{eqnarray}
{\mathcal U} &=& \frac{1}{8 a^2 \ell_{\rm P}^2}\biggl[ 5 \,H_{k\ell}\,H^{k\ell} - 7
\partial_{m} h_{k\ell} \partial^{m} h^{k\ell} \biggr],
\label{LL3}\\
\Pi_{i}^{j} &=& 
\frac{1}{a^2 \ell_{\rm P}^2} \biggl\{ \frac{1}{6} \biggl[ H_{k\ell} H^{k\ell} - 
\frac{1}{2} \partial_{m} h_{k\ell} \partial^{m} h^{k\ell} \biggr] \delta_{i}^{j}
+ \frac{1}{2} \partial_{m} h_{\ell i} \partial^{m} h^{\ell j} 
\nonumber\\
&-& \frac{1}{4} \partial_{i} h_{k\ell} \partial^{j}  h^{k\ell} 
- \frac{1}{2} H_{k i}\, H^{k j} \biggr\},
\label{LL4}
\end{eqnarray}
 with $\Pi_{i}^{i} =0$. In Eqs. (\ref{LL1})--(\ref{LL2}) and (\ref{LL3})--(\ref{LL4}) we used the notation 
 \begin{equation}
 \partial_{\tau} h_{i j} = h_{i j}^{\prime}= H_{ij}, \qquad {\mathcal H} = \frac{a^{\prime}}{a}. 
 \label{DEFNOT}
 \end{equation}
 
 In a complementary perspective, from Eq. (\ref{FPdef2}), the  energy momentum pseudo-tensor can be expressed as: 
 \begin{equation}
{\mathcal W}_{\mu}^{\nu} = \frac{1}{4 \ell_{\mathrm{P}}^2 a^2} \biggl[ \partial_{\mu} h_{i j} \partial^{\nu} h^{i j} 
- \frac{1}{2} \delta_{\mu}^{\nu} \,\overline{g}^{\alpha\beta}\, \partial_{\alpha} h_{ij} \partial_{\beta} h^{ij} \biggr],
\label{FF1}
\end{equation}
or, in components, 
\begin{eqnarray}
{\mathcal W}_{0}^{0} &=& \frac{1}{8 \ell_{\mathrm{P}}^2 a^2} \biggl[ H_{k \ell}\, H^{k \ell} + \partial_{m} h_{k\ell} \partial_{m} h^{k\ell}\biggr],
\label{FF2}\\ 
{\mathcal W}_{i}^{j} &=& \frac{{\mathcal W}}{3} \delta_{i}^{j} + \Pi_{i}^{j},\qquad {\mathcal W}_{i}^{0} =  \frac{1}{4 \ell_{\mathrm{P}}^2 a^2} H_{k\ell} \,\partial_{i} h^{k\ell},
\label{FF3}
\end{eqnarray}
where $\Pi_{i}^{i} =0$ and, moreover, 
\begin{eqnarray}
{\mathcal W} &=&  \frac{1}{8 \ell_{\mathrm{P}}^2 a^2} \biggl[ \partial_{m} h_{k \ell} \partial_{m} h^{k\ell} - 3 H_{k\ell}
h^{k\ell} \biggr],
\label{FF4}\\
\Pi_{i}^{j} &=& \frac{1}{4 \ell_{\mathrm{P}}^2 a^2} \biggl[ - \partial_{i} h_{k\ell} \partial^{j} h^{k\ell} + \frac{1}{3} \delta_{i}^{j} \partial_{m} h_{k \ell} \partial_{m} h_{k\ell} \biggr].
\label{FF5}
\end{eqnarray}
\subsection{Polarizations of the tensor modes}
With the aim of discussing more explicit expression of ${\mathcal U}_{\mu}^{\nu}$ and ${\mathcal W}_{\mu}^{\nu}$ it is 
useful to introduce the two polarizations of the gravitons:
 \begin{equation}
 e_{ij}^{(\oplus)}(\hat{k}) = (\hat{m}_{i} \hat{m}_{j} - \hat{n}_{i} \hat{n}_{j}), \qquad 
 e_{ij}^{(\otimes)}(\hat{k}) = (\hat{m}_{i} \hat{n}_{j} + \hat{n}_{i} \hat{m}_{j}),
 \label{ST0}
 \end{equation}
 where $\hat{k}$ is oriented along the direction of propagation of the wave while $\hat{m}_{i} = m_{i}/|\vec{m}|$ and $\hat{n} =n_{i}/|\vec{n}|$ are mutually
 orthogonal and orthogonal to $\hat{k}$. 
  It follows from Eq. (\ref{ST0}) that $e_{ij}^{(\lambda)}\,e_{ij}^{(\lambda')} = 2 \delta_{\lambda\lambda'}$ while the sum over the polarizations gives:
\begin{equation}
\sum_{\lambda} e_{ij}^{(\lambda)}(\hat{k}) \, e_{m n}^{(\lambda)}(\hat{k}) = \biggl[p_{m i}(\hat{k}) p_{n j}(\hat{k}) + p_{m j}(\hat{k}) p_{n i}(\hat{k}) - p_{i j}(\hat{k}) p_{m n}(\hat{k}) \biggr];
\label{ST0B} 
\end{equation}
with $p_{ij}(\hat{k}) = (\delta_{i j} - \hat{k}_{i} \hat{k}_{j})$. The Fourier transform of $h_{ij}(\vec{x},\tau)$ is defined as:
\begin{equation}
h_{ij}(\vec{k},\tau) = \frac{1}{(2\pi)^{3/2}}\, \sum_{\lambda}  \, \int d^{3} k \, h_{ij}(\vec{x},\tau)\,\, e^{ i \vec{k}\cdot\vec{x}}.
\label{St0C}
\end{equation}
In the quantum description of the evolution of the gravitons $h_{ij}(\vec{x},\tau)$ and $H_{ij}(\vec{x},\tau)$  interpreted as field operators in the Heisenberg  description, namely:
\begin{eqnarray}
\hat{h}_{ij}(\vec{x},\tau) &=& \frac{\sqrt{2} \ell_{\mathrm{P}}}{(2\pi)^{3/2} a}\sum_{\lambda} \int \, d^{3} k \,\,e^{(\lambda)}_{ij}(\vec{k})\, [ f_{k,\lambda}(\tau) \hat{a}_{\vec{k}\,\lambda } e^{- i \vec{k} \cdot \vec{x}} + f^{*}_{k,\lambda}(\tau) \hat{a}_{\vec{k}\,\lambda }^{\dagger} e^{ i \vec{k} \cdot \vec{x}} ],
\label{T8}\\
\hat{H}_{ij}(\vec{x},\tau) &=& \frac{\sqrt{2} \ell_{\mathrm{P}}}{(2\pi)^{3/2} a}\sum_{\lambda} \int \, d^{3} k \,\,e^{(\lambda)}_{ij}(\vec{k})\, [ g_{k,\lambda}(\tau) \hat{a}_{\vec{k}\,\lambda } e^{- i \vec{k} \cdot \vec{x}} + g^{*}_{k,\lambda}(\tau) \hat{a}_{\vec{k}\,\lambda }^{\dagger} e^{ i \vec{k} \cdot \vec{x}} ],
\label{T8a}
\end{eqnarray}
where $f_{k}$ and $g_{k} = (f_{k}' - {\mathcal H} f_{k})$ are the corresponding mode functions; the creation and annihilation operators 
obey, as usual, $[\hat{a}_{\vec{k},\,\lambda}, \, \hat{a}^{\dagger}_{\vec{k}^{\prime},\,\lambda^{\prime}}]  = \delta_{\lambda\,\lambda^{\prime}} \, \delta^{(3)}(\vec{k}- \vec{k}^{\prime})$. In the parametrization (\ref{St0C})  
where $\hat{h}_{ij}(\vec{k},\tau)$ and $\hat{H}_{ij}(\vec{k},\tau)$ are given by:
\begin{eqnarray}
\hat{h}_{ij}(\vec{k},\tau) &=& \frac{\sqrt{2} \ell_{\mathrm{P}}}{a} \sum_{\lambda} \biggl[ e^{(\lambda)}_{ij}(\vec{k})\, f_{k,\lambda}(\tau) \hat{a}_{\vec{k}\,\lambda } + e^{(\lambda)}_{ij}(-\vec{k})\, f^{*}_{k,\lambda}(\tau) \hat{a}_{-\vec{k}\,\lambda }\biggr],
\nonumber\\
\hat{H}_{ij}(\vec{k},\tau) &=& \frac{\sqrt{2} \ell_{\mathrm{P}}}{a} \sum_{\lambda} \biggl[ e^{(\lambda)}_{ij}(\vec{k})\, g_{k,\lambda}(\tau) \hat{a}_{\vec{k}\,\lambda } + e^{(\lambda)}_{ij}(-\vec{k})\, g^{*}_{k,\lambda}(\tau) \hat{a}_{-\vec{k}\,\lambda }\biggr].
\label{T8b}
\end{eqnarray}

From Eq. (\ref{T8b}) the two-point functions of the operators can be expressed in terms of the related power spectra:
\begin{eqnarray}
\langle \hat{h}_{ij}(\vec{k},\tau) \, \hat{h}_{mn}(\vec{k}^{\prime}, \tau) \rangle &=& \frac{2 \pi^2}{k^3}\, P_{ff}(k,\tau) \, \delta^{(3)}(\vec{k} + \vec{k}^{\prime}) \, 
{\mathcal S}_{i j m n}(\hat{k}),
\nonumber\\
\langle \hat{H}_{ij}(\vec{k},\tau) \, \hat{H}_{mn}(\vec{k}^{\prime}, \tau) \rangle &=& \frac{2 \pi^2}{k^3}\, P_{gg}(k,\tau) \, \delta^{(3)}(\vec{k} + \vec{k}^{\prime}) \, 
{\mathcal S}_{i j m n}(\hat{k}),
\nonumber\\
\langle \hat{h}_{ij}(\vec{k},\tau) \, \hat{H}_{mn}(\vec{k}^{\prime}, \tau) \rangle &=& \frac{2 \pi^2}{k^3}\, P_{fg}(k,\tau) \, \delta^{(3)}(\vec{k} + \vec{k}^{\prime}) \, 
{\mathcal S}_{i j m n}(\hat{k}),
\nonumber\\
\langle \hat{H}_{ij}(\vec{k},\tau) \, \hat{h}_{mn}(\vec{k}^{\prime}, \tau) \rangle &=& \frac{2 \pi^2}{k^3}\, P_{gf}(k,\tau) \, \delta^{(3)}(\vec{k} + \vec{k}^{\prime}) \, 
{\mathcal S}_{i j m n}(\hat{k}).
\label{PS}
\end{eqnarray}
The power spectra appearing in Eq. (\ref{PS}) are expressible in terms of the corresponding mode functions:
\begin{eqnarray}
P_{ff}(k,\tau) &=&  \frac{4 \ell_{\mathrm{P}}^2\,\, k^3}{\pi^2\, a^2(\tau)} |f_{k}(\tau)|^2, \qquad P_{gg}(k,\tau) =  \frac{4 \ell_{\mathrm{P}}^2\,\, k^3}{\pi^2\, a^2(\tau)} |g_{k}(\tau)|^2,
\label{FFGG}\\
P_{fg}(k,\tau) &=&  \frac{4 \ell_{\mathrm{P}}^2\,\, k^3}{\pi^2\, a^2(\tau)} f_{k}(\tau) g_{k}^{\ast}(\tau),\qquad P_{gf}(k,\tau) =  \frac{4 \ell_{\mathrm{P}}^2\,\, k^3}{\pi^2\, a^2(\tau)} f^{\ast}_{k}(\tau) g_{k}(\tau).
\label{FGGF}
\end{eqnarray}
In Eq. (\ref{PS}) we introduced the following quantity: 
\begin{equation}
{\mathcal S}_{ijmn}(\hat{k}) = \frac{1}{4} \sum_{\lambda} e_{ij}^{(\lambda)}(\hat{k}) \, e_{m n}^{(\lambda)}(\hat{k}).
\label{ST1B}
\end{equation}
Recalling Eqs. (\ref{ST0}) and (\ref{ST0B}), it is easy to derive the following identity:
\begin{equation}
{\mathcal Q}(\vec{a}, \vec{b}) = {\mathcal S}_{i j m n}(\hat{a}) {\mathcal S}_{i j m n}(\hat{b}) = \frac{[ 1 + (\hat{a}\cdot\hat{b})^2] [ 1 + 3  (\hat{a}\cdot\hat{b})^2]}{16},
\label{ID1}
\end{equation}
where $\hat{a} = \vec{a}/|\vec{a}|$ and $\hat{b} = \vec{b}/|\vec{b}|$. If $\hat{a}$ and $\hat{b}$ coincide ${\mathcal Q}(\hat{a},\hat{a}) \to 1/2$.

\renewcommand{\theequation}{3.\arabic{equation}}
\setcounter{equation}{0}
\section{The scalar modes of the relic graviton fluid}
\label{sec3}
\subsection{Governing equations}
In the long wavelength limit (i.e. when $k\tau \ll 1$)  the modes of the field are amplified and the fluctuations of the pressure and of the energy density 
depend on the salient features of the background evolution. 
In the opposite regime, i.e. $k\tau \gg 1 $ (short wavelength limit) the Fourier modes of the tensor 
fluctuations of the geometry oscillate. From the discussion of section \ref{sec4} the perturbed entries 
are always the same but their relation with the amplitudes of the two tensor polarizations depends on the 
specific parametrization (be it ${\mathcal W}_{\mu}^{\nu}$ or ${\mathcal U}_{\mu}^{\nu}$).
For the purposes of the present section it is sufficient to posit the following general form\footnote{Recalling the considerations of section \ref{sec2}, ${\mathcal T}_{\mu}^{\nu}$ may either coincide with ${\mathcal U}_{\mu}^{\nu}$ or with ${\mathcal W}_{\mu}^{\nu}$; 
note that ${\mathcal J}_{i}$ (which turns out to be subleading in the long wavelength approximation) is the same both in the Landau-Lifshitz and in the Ford-Parker parametrizations.} for the components  of $\delta_{\mathrm{s}} {\mathcal T}_{\mu}^{\nu}$:
\begin{eqnarray}
\delta_{\mathrm{s}} {\mathcal T}_{0}^{0} = \delta \rho_{\mathrm{gw}}, \qquad \delta_{\mathrm{s}} {\mathcal T}_{i}^{0} = {\mathcal J}_{i} =
 \frac{ H_{k\ell}\,  \partial_{i} h^{k \ell}}{4 \ell_{\mathrm{P}}^2 a^2}
,\qquad 
\delta_{\mathrm{s}} {\mathcal T}_{i}^{j} &=&- \delta p_{\mathrm{gw}} \delta_{i}^{j} + \Pi_{i}^{j}.
\label{PRAM}
\end{eqnarray}
The fluctuations of the energy density and pressure are:
\begin{equation}
\rho_{\mathrm{gw}}(\vec{x},\tau) = \langle \rho_{\mathrm{gw}}\rangle + \delta \rho_{\mathrm{gw}}, \qquad p_{\mathrm{gw}}(\vec{x},\tau) = \langle p_{\mathrm{gw}}\rangle + \delta p_{\mathrm{gw}},
\label{defluc}
\end{equation}
where $\langle .\,.\,.\,.\rangle$ denotes the expectation value on the quantum state 
of the relic gravitons\footnote{Note that, in spite of the parametrization, $\langle \Pi_{ij} \rangle =0$ since the anisotropic stress is fully inhomogeneous.}; the expectation values can be also viewed as stochastic averages by using 
the correspondence between the power spectra and the mode functions 
outlined in Eq. (\ref{PS}). Different forms pseudo-tensors (see e.g. Eqs. (\ref{EIN6}) and (\ref{FF1})) modify the actual expressions of
$\delta \rho_{\mathrm{gw}}$, $ \delta p_{\mathrm{gw}}$ and  $\Pi_{i}^{j}$. In section \ref{sec4} this 
point will be examined in detail but the relation between the induced curvature perturbations and Eq. (\ref{PRAM})
 that are independent on the specific parametrization. 

In the sudden reheating approximation,  the quasi-de Sitter stage of expansion lasts up to a fiducial time
$\tau_{1}$ marking simultaneously the end of inflation and the onset 
of radiation. In the short wavelength regime we have a collection 
of relativistic bosons; thus the pressure fluctuations  will be given by $\delta p_{\mathrm{gw}} = \delta \rho_{\mathrm{gw}}/3$. 
This implies that the short wavelengths fluctuations of the energy density and of the pressure are exponentially suppressed during the 
de Sitter stage of expansion while the background energy density is nearly constant. 

Conversely the inhomogeneities induced by the long wavelength gravitons may affect the curvature perturbations 
after inflation: during the radiation-dominated phase the long wavelength modes 
decrease more slowly than the background energy density (going as $a^{-4}$ in the conventional post-inflationary thermal history).
With these specifications, and recalling that  $h_{ij}$ is  gauge-invariant for infinitesimal coordinate 
transformations preserving the tensor nature of the fluctuation, we can consider the scalar modes of the geometry in the presence of the fluctuations of the energy density 
and pressure induced by the relic gravitons.  Using Eq. (\ref{PRAM}),  the $(00)$ and $(0i)$ components of Eq. (\ref{EIN5}) give the Hamiltonian and the momentum constraints
whose explicit form is given by:
\begin{eqnarray}
&& \nabla^2 \Psi - 3 {\cal H} ( \Psi' + {\cal H} \Phi) = \frac{a^2 \ell_{\mathrm{P}}^2}{2} [ \delta \rho_{\mathrm{t}} + \delta\rho_{\mathrm{gw}}],
\label{GS1}\\
&& \nabla^2( \Psi' + {\cal H} \Phi) = -\frac{a^2  \ell_{\mathrm{P}}^2}{2}[ (p_{\mathrm{t}} + \rho_{\mathrm{t}}) \theta_{\mathrm{t}}+ \vec{\nabla} \cdot \vec{{\mathcal J}}],
\label{GS2}
\end{eqnarray}
where $\theta_{\mathrm{t}}$ is, as usual, the three-divergence of the total velocity field while $\Phi$ and $\Psi$ are the gauge-invariant Bardeen potentials \cite{bard1};
$\delta \rho_{\mathrm{t}}$ and $\delta p_{\mathrm{t}}$  denote the scalar fluctuations of the total energy density and total pressure of the fluid 
during the post-inflationary phase. The $(i=j)$ 
and $(i\neq j)$ components of Eq. (\ref{EIN5}) read instead:
\begin{eqnarray}
&& \Psi'' + {\cal H} ( \Phi' + 2 \Psi') + ( {\cal H}^2 + 2 {\cal H}') \Phi  + \frac{1}{3} \nabla^2 ( \Phi - \Psi) = \frac{a^2 \ell_{\mathrm{P}}}{2} (\delta p_{\mathrm{t}} + \delta p_{\mathrm{gw}}),
\label{GS3}\\
&& \nabla^2 ( \Phi - \Psi) =  \frac{3}{2} \ell_{\mathrm{P}}^2 a^2 \Pi_{\mathrm{gw}},
\label{GS4} 
\end{eqnarray}
where the following practical notation: 
\begin{equation}
 \partial_{i}\partial_{j} \Pi^{ij} = \nabla^2 \Pi_{\mathrm{gw}}
\label{GS4a}
\end{equation}
has been introduced. In connection with the anisotropic stress we remark that all sources of anisotropic stress besides the relic gravitons have been 
neglected but this will not alter the conclusions of the analysis.

\subsection{Curvature perturbations}
The evolution of the curvature perturbations can be easily obtained by introducing the total sound 
speed of the system, i.e. $c_{\mathrm{st}}^2 = p_{\mathrm{t}}^{\prime}/\rho_{\mathrm{t}}^{\prime}$.
Multiplying Eq. (\ref{GS1}) by $c_{\mathrm{st}}^2$ and subtracting the obtained result from 
Eq. (\ref{GS3}) we get to the relation:
\begin{eqnarray}
&& \Psi'' + {\mathcal H}[ \Phi' + ( 2 + 3 c_{\rm st}^2 ) \Psi'] + 
[ {\mathcal H}^2 ( 1 + 3 c_{\rm st}^2) + 2 {\mathcal H}'] \Phi 
\nonumber\\
&& - c_{\rm st}^2 \nabla^2 \Psi + \frac{1}{3}\nabla^2( \Phi - \Psi) =
\frac{a^2\ell_{\mathrm{P}}^2}{2}[ \delta p_{\mathrm{nad}} + \delta p_{\mathrm{gw}} - c_{\mathrm{st}}^2 \delta\rho_{\mathrm{gw}} ].
\label{GS5}
\end{eqnarray}
The standard notation $\delta p_{\mathrm{nad}} = \delta p_{\mathrm{t}} - c_{\mathrm{st}}^2 \delta\rho_{\mathrm{t}}$ has been introduced in Eq. (\ref{GS5}). In the 
conventional adiabatic scenario (where non-adiabatic modes are, by definition,  absent) we have $\delta p_{\mathrm{nad}} =0$.

From the gauge-invariant expression of the curvature perturbations\footnote{We recall that ${\mathcal R}$ is defined as 
${\mathcal R} = - \Psi - {\mathcal H}({\mathcal H} \Phi + \Psi')/({\mathcal H}^2 - {\mathcal H}')$.} 
on comoving orthogonal hypersurfaces \cite{bard1} (see also \cite{lukash}) we obtain the relation:
\begin{eqnarray}
{\mathcal R}' &=& \Sigma_{{\mathcal R}} - \frac{2 {\mathcal H}\, c_{\mathrm{st}}^2 \, \nabla^2 \Psi}{\ell_{\mathrm{P}}^2 a^2 (\rho_{\mathrm{t}} + p_{\mathrm{t}})},
\label{GS6}\\
\Sigma_{{\mathcal R}} &=& - \frac{{\mathcal H}}{p_{\mathrm{t}} + \rho_{\mathrm{t}}} \delta p_{\mathrm{nad}} +
\frac{{\mathcal H}}{p_{\mathrm{t}} + \rho_{\mathrm{t}}}\biggl[ c_{\mathrm{st}}^2 \delta\rho_{\mathrm{gw}} - \delta p_{\mathrm{gw}} + \Pi_{\mathrm{gw}} \biggr].
\label{GS7}
\end{eqnarray}
Equation (\ref{GS6}) is not decoupled since it depends on $\Psi$. By taking the first derivative of both sides 
of Eq. (\ref{GS6}) and by using Eq. (\ref{GS4}) in the obtained expression we obtain the following evolution equation:
\begin{equation}
{\mathcal R}'' + 2 \frac{z_{\mathrm{t}}'}{z_{\mathrm{t}}} {\mathcal R}' - c_{\mathrm{st}}^2 \nabla^2 {\mathcal R} = 
\Sigma_{{\mathcal R}}' +  2 \frac{z_{\mathrm{t}}'}{z_{\mathrm{t}}} \Sigma_{{\mathcal R}} + \frac{3 a^4}{z_{\mathrm{t}}^2} \Pi_{\mathrm{gw}}, 
\label{GS8}
\end{equation}
where $z_{\mathrm{t}}$ is defined as $z_{\mathrm{t}} = a^2 \sqrt{p_{\mathrm{t}} + \rho_{\mathrm{t}}}/( c_{\mathrm{st}} {\mathcal H})$.
To derive Eqs. (\ref{GS8}) we need to use the evolution equations of the background written in the 
conformally flat case, i.e. 
\begin{equation}
3 {\mathcal H}^2 = \ell_{\mathrm{P}}^2 a^2 \rho_{t}, \qquad 2({\mathcal H}^2 - {\mathcal H}' )= \ell_{\mathrm{P}}^2 a^2 (\rho_{t} + p_{t});
\label{GS9a}
\end{equation}
combining Eqs. (\ref{GS9a}) we get, as expected, the result coming from the covariant conservation 
of the energy-momentum tensor of the sources, i.e. $\rho_{t}' + 3 {\mathcal H} (\rho_{t} + p_{t}) =0$. 
We must remind that the possible effect of the energy density of the gravitons on the evolution of the 
background has been totally neglected due to its smallness. Relic gravitons may instead contribute
to the evolution of the curvature perturbations as Eq. (\ref{GS8}) shows.

\subsection{Large-scale solutions with relic gravitons}
Equation  (\ref{GS8}) will now be studied for typical wavelengths larger than the Hubble radius. In this case 
the term $c_{\mathrm{st}}^2 \nabla^2 {\mathcal R}$ can be consistently dropped but
it is appropriate to stress that it would be incorrect to eliminate the contribution of the gradients directly in Eq. (\ref{GS6}). By dropping $\nabla^2 \Psi$ 
we would also neglect terms that are comparable to the contribution of the anisotropic stress of the relic gravitons. We could 
neglect the presence of any non-adiabatic mode in the fluid sector by setting $\delta p_{\mathrm{nad}} =0$ in Eq. (\ref{GS7}). However this 
assumption is not necessary insofar as we will write down a general solution containing also the first derivative of the curvature perturbations.

With the previous specifications, Eq. (\ref{GS8}) can be first written in the form of an integral equation as 
\begin{eqnarray}
&&{\mathcal R}(\vec{x},\tau) = {\mathcal R}_{\ast}(\vec{x}) + z_{\mathrm{t}}^2(\tau_{1}) ({\mathcal R}' - \Sigma_{{\mathcal R}})_{\tau_{1}} \, \int_{\tau_{1}}^{\tau} \frac{ d \tau'}{z_{\mathrm{t}}^2(\tau')} 
+ \int_{\tau_{1}}^{\tau} \Sigma_{{\mathcal R}}(\vec{x},\tau') \, \, d\tau' 
\nonumber\\
&+& 3 \int_{\tau_{1}}^{\tau} \frac{ d \tau'}{z_{\mathrm{t}}^2(\tau')} \int_{\tau_{1}}^{\tau'} a^4(\tau^{\prime\prime}) \, \Pi_{\mathrm{gw}}(\vec{x},\tau^{\prime\prime}) \, d\tau^{\prime\prime}+ \int_{\tau_{1}}^{\tau} \frac{d\tau'}{z_{\mathrm{t}}^2(\tau^{\prime})} \, \int_{\tau_{1}}^{\tau^{\prime}} \,z_{\mathrm{t}}^2(\tau^{\prime\prime}) c_{\mathrm{st}}^2(\tau^{\prime\prime}) \nabla^2{\mathcal R}(\vec{x},\tau^{\prime\prime}) \, d\tau^{\prime\prime}.
\nonumber
\end{eqnarray}
that can be formally solved by iteration in the long wavelength limit where the term $ c_{\mathrm{st}}^2\,\nabla^2 {\mathcal R}_{*}$ can be 
neglected in comparison with the other contributions. Thus the long wavelength solution of Eq. (\ref{GS8}) can be written, 
\begin{eqnarray}
{\mathcal R}(\vec{x},\tau) &=& {\mathcal R}_{\ast}(\vec{x}) + z_{\mathrm{t}}^2(\tau_{1}) ({\mathcal R}' - \Sigma_{{\mathcal R}})_{\tau_{1}} \, \int_{\tau_{1}}^{\tau} \frac{ d \tau'}{z_{\mathrm{t}}^2(\tau')} 
\nonumber\\
&+& \int_{\tau_{1}}^{\tau} \Sigma_{{\mathcal R}}(\vec{x},\tau') \, \, d\tau' 
\nonumber\\
&+& 3 \int_{\tau_{1}}^{\tau} \frac{ d \tau'}{z_{\mathrm{t}}^2(\tau')} \int_{\tau_{1}}^{\tau'} a^4(\tau^{\prime\prime}) \, \Pi_{\mathrm{gw}}(\vec{x},\tau^{\prime\prime}) \, d\tau^{\prime\prime} + {\mathcal O} (c_{\mathrm{st}}^2\,\nabla^2 {\mathcal R}_{*}).
\label{GS10}
\end{eqnarray}
The solution reported in Eq. (\ref{GS10}) has been written in general without assuming anything on the 
adiabatic (or non-adiabatic) nature of the dominant source of inhomogeneity. In what follows we shall focus however to 
 the case where $\delta p_{\mathrm{and}} =0$.  Note that Eq. (\ref{GS10}) is written in real space but it can be 
 easily translated in Fourier space. Before proceeding further we need to compute the evolution of  the various terms appearing inside the integrals.
 These will be the main themes of the forthcoming section \ref{sec4}.

\renewcommand{\theequation}{4.\arabic{equation}}
\setcounter{equation}{0}
\section{Power spectra and explicit estimates}
\label{sec4}
The calculation of the various power 
spectra  present similar technical aspects so that we shall discuss 
the case of the energy density and then mention the salient results
for the pressure and the anisotropic stress. To avoid digressions a number of 
explicit results have been relegated to the appendices 
\ref{appAA}, \ref{appA} and \ref{appB}.

\subsection{Power spectrum of the energy density of the relic gravitons}
Let us therefore go back to the fluctuations of the energy density defined in Eq. (\ref{defluc}) and introduce their Fourier transform:
\begin{equation}
\delta\rho_{\mathrm{gw}}(\vec{x},\tau) = \frac{1}{(2\pi)^{3/2}} \int d^{3} k \, e^{- i \vec{q} \cdot \vec{x}}\, \delta\rho_{\mathrm{gw}}(\vec{q},\tau),
\label{EN1}
\end{equation}
whose corresponding power spectrum is defined, within the present conventions, as:
\begin{equation}
\langle \delta \rho_{\mathrm{gw}}(\vec{p},\tau) \delta \rho_{\mathrm{gw}}(\vec{q},\tau) \rangle = \frac{2\pi^2}{q^3}\, 
\delta^{(3)}(\vec{q} + \vec{p}) \, {\mathcal P}_{\rho_{\mathrm{gw}}}(q,\tau).
\label{EN2}
\end{equation} 
According to the definition (\ref{defluc}), $\langle \delta\rho_{\mathrm{gw}}(\vec{q},\tau) \rangle =0$; moreover
 Eq. (\ref{FF2})  
(in the Ford-Parker parametrization) and Eq. (\ref{LL1}) (in the Landau-Lifshitz parametrization) lead, in Fourier space, to the following results:
\begin{eqnarray}
&& \delta \rho^{(FP)}_{\mathrm{gw}}(\vec{q},\tau) = \frac{1}{8 \ell_{\mathrm{P}}^2 a^2 } \int \frac{d^{3} k}{(2\pi)^{3/2}} \biggl\{ \hat{H}_{\ell m}(\vec{k},\tau) 
\hat{H}_{\ell m}(\vec{q} -\vec{k},\tau)
\nonumber\\
&&+ [\vec{k}\cdot(\vec{k} - \vec{q})]\, \hat{h}_{\ell m}(\vec{k},\tau) 
 \hat{h}_{\ell m}(\vec{q} -\vec{k},\tau) 
- \frac{2\pi^2}{k^3} \biggl[ {\mathcal P}_{gg}(k,\tau) + k^2 {\mathcal P}_{ff}(k,\tau) \biggr] \delta^{(3)}(\vec{q}) \biggr\},
 \label{EN3}\\
&& \delta \rho^{(LL)}_{\mathrm{gw}}(\vec{q},\tau) =\frac{1}{8 \ell_{\mathrm{P}}^2 a^2} \int \frac{d^{3} k}{(2\pi)^{3/2}} \biggl\{ \hat{H}_{\ell m}(\vec{k},\tau) 
 \hat{H}_{\ell m}(\vec{q} -\vec{k},\tau)
 \nonumber\\
&&+   [\vec{k}\cdot(\vec{k} - \vec{q})] \, \hat{h}_{\ell m}(\vec{k},\tau) 
 \hat{h}_{\ell m}(\vec{q} -\vec{k},\tau) 
\nonumber\\
&&+ 4 {\mathcal H} \biggl[ \hat{H}_{\ell m}(\vec{k},\tau) 
 \hat{h}_{\ell m}(\vec{q} -\vec{k},\tau) + \hat{h}_{\ell m}(\vec{k},\tau) 
 \hat{H}_{\ell m}(\vec{q} -\vec{k},\tau)\biggr]
\nonumber\\
&& - \frac{2\pi^2}{k^3} \biggl[ {\mathcal P}_{gg}(k,\tau) + k^2 {\mathcal P}_{ff}(k,\tau) + 4 {\mathcal H} \biggl( {\mathcal P}_{fg}(k,\tau) + {\mathcal P}_{gf}(k,\tau) \biggl)\biggr]  \delta^{(3)}(\vec{q})
 \biggr\},
\label{EN4}
\end{eqnarray}
where, according to Eq. (\ref{DEFNOT}), the notation 
$\hat{H}_{\ell m}(\vec{k},\tau) =  \hat{h}^{\prime}_{\ell m}(\vec{k},\tau)$ has been introduced. 

The power spectra of the fluctuation defined in Eq. (\ref{EN2}) can be written, 
with the help of Eqs. (\ref{EN3}) and (\ref{EN4}), as:
\begin{eqnarray}
&&{\mathcal P}_{\rho_{\mathrm{gw}}}^{(FP)}(q,\tau) = \frac{q^3}{128\pi \ell_{\mathrm{P}}^4 a^4(\tau)} 
\int \frac{d^{3}k}{k^3 \, |\vec{q} - \vec{k}|^3} \, {\mathcal Q}(\vec{k},\vec{q})\,\biggl\{ {\mathcal P}_{gg}(k,\tau)\, {\mathcal P}_{gg}(|\vec{q} - \vec{k}|,\tau)
\nonumber\\
&& +\vec{k}\cdot(\vec{q} - \vec{k}) \, \biggl[ {\mathcal P}_{fg}(k,\tau) {\mathcal P}_{gf}(|\vec{q} - \vec{k}|,\tau)+ {\mathcal P}_{gf}(k,\tau) {\mathcal P}_{fg}(|\vec{q} - \vec{k}|,\tau)\biggr]
\nonumber\\
&&+ [ \vec{k}\cdot(\vec{q} - \vec{k})]^2 {\mathcal P}_{ff}(k,\tau)\, {\mathcal P}_{ff}(|\vec{q} - \vec{k}|,\tau)\biggl\},
\label{EN6}\\
&& {\mathcal P}_{\rho_{\mathrm{gw}}}^{(LL)}(q,\tau) = \frac{q^3}{128\pi \ell_{\mathrm{P}}^4 a^4(\tau)} 
\int \frac{d^{3}k}{k^3 \, |\vec{q} - \vec{k}|^3} \, {\mathcal Q}(\vec{k},\vec{q})\,\biggl[ 16 \,{\mathcal H}^2\, {\mathcal A}_{1}(\vec{k}, \vec{q})
\nonumber\\
&&+ 4\, {\mathcal H}\, {\mathcal A}_{2}(\vec{k}, \vec{q}) + {\mathcal A}_{3}(\vec{k}, \vec{q})\biggr];
\label{ENN8}
\end{eqnarray}
the functions  ${\mathcal A}_{1}(\vec{k},\vec{q},\tau)$, ${\mathcal A}_{2}(\vec{k},\vec{q},\tau)$ and 
${\mathcal A}_{3}(\vec{k},\vec{q},\tau)$ are defined in Eqs. (\ref{ENN9}), (\ref{ENN10}) and (\ref{ENN11}) of appendix \ref{appAA};
note that ${\mathcal Q}(\vec{k},\vec{q})$ can be computed thanks to the identity 
given in Eq. (\ref{ID1}) by identifying $\hat{a} = \hat{k}$ and $\hat{b}=(\vec{q} - \vec{k})/|\vec{q} - \vec{k}|$:
\begin{equation}
{\mathcal Q}(\vec{k},\vec{q}) = \frac{1}{16} \biggl[ 1 + 4 \frac{(\vec{k}\cdot\vec{q} - k^2)^2}{k^2 |\vec{q} - \vec{k}|^2} + 3 \frac{(\vec{k}\cdot\vec{q} - k^2)^4}{k^4 |\vec{q} - \vec{k}|^4}\biggr].
\label{ENN7}
\end{equation}
The presence of further terms in Eq. (\ref{EN4}) complicates the correlator and it is ultimately responsible of the form of the power spectrum 
of Eq. (\ref{ENN8}). Thus the differences between Eqs. (\ref{EN6}) and (\ref{ENN8}) are a direct consequence of the form of  Eqs. (\ref{EN3}) and (\ref{EN4}).  We remark that Eqs. (\ref{EN6}) and (\ref{ENN8}) have been written 
in general  but $P_{ff}(k,\tau)$, $P_{gg}(k,\tau)$, 
$P_{fg}(k,\tau)$ and $P_{gf}(k,\tau)$  depend on the mode functions according to Eqs. (\ref{FFGG}) and (\ref{FGGF}). 
The mode functions are determined, in their turn, by the evolution of the background that will now be specified. 

\subsection{Explicit forms for the mode functions}
We shall be dealing hereunder with the following analytical form of the scale factor\footnote{Equations (\ref{dsa}) and (\ref{rada}) are continuous (with the first derivatives) in $\tau = - \tau_{1}$. 
Similarly Eqs. (\ref{rada}) and (\ref{radmat}) are continuous (with their first derivatives) in $\tau= \tau_{2}$.}
\begin{eqnarray}
a(\tau) &=& - \frac{\tau_{1}}{\tau},\qquad \tau \leq - \tau_{1},
\label{dsa}\\
a(\tau) &=& \frac{\tau + 2 \tau_{1}}{\tau_{1}}, \qquad - \tau_{1} < \tau < \tau_{2},
\label{rada}\\
a(\tau) &=& \frac{( \tau + \tau_{2} + 4 \tau_{1})^2}{4 \tau_{1} ( \tau_{2} + 2 \tau_{1})}, \qquad \tau > \tau_{2}.
\label{radmat} 
\end{eqnarray}
The typical time-scale $\tau_{1}$  denotes  the transition from 
the de Sitter stage of expansion to the radiation-dominated epoch ending at the conformal time $\tau_{2}$ which coincides 
with the equality time and defines the beginning of the matter-dominated period.
of expansion. 

Equation (\ref{dsa}) implies that  the mode function solving Eq. (\ref{MFUN}) for $\tau \leq -\tau_1$  is:
\begin{equation}
f_{k}(\tau) = \frac{1}{\sqrt{2 k}} \biggl( 1 - \frac{i}{x}\biggr)\,\, e^{- i x},\qquad
f^{\prime}_{k}(\tau) = i \sqrt{\frac{k}{2}} 
\biggl[ \frac{1}{x^2} - 1 + \frac{i}{x}\biggr]\,\,e^{- i x},
\label{modf}
\end{equation}
where we introduced the dimensionless combination $ x = k\,\tau$. For $\tau >- \tau_{1}$ the continuity of the canonical field operators demands 
the continuity of the mode functions, i.e. 
\begin{eqnarray}
f_{k}(-\tau_1) = c_{+}(k) F_{k}(-\tau_1) + c_{-}(k) F_{k}^{*}(-\tau_1),
\nonumber\\
f'_{k}(-\tau_1) = c_{+}(k) F'_{k}(-\tau_1) + c_{-}(k) F_{k}^{\prime\,*}(-\tau_1),
\label{BogB}
\end{eqnarray}
where $F_{k}(\tau)$ and $ F^{\prime}_{k}(\tau) $ are now given by:
\begin{equation}
F_{k}(\tau) = \frac{1}{\sqrt{2 k} } e^{ - i  (x + 2 x_1)},\qquad
  F^{\prime}_{k}(\tau) = -i \sqrt{\frac{k}{2}} e^{- i ( x + 2 x_1)},
\label{modr3}
\end{equation}
and solve Eq. (\ref{MFUN}) during the radiation-dominated stage (see Eq. (\ref{rada})).
Inserting Eqs. (\ref{modf}) and (\ref{modr3}) into Eq. (\ref{BogB}),  $c_{\pm}(x_{1})$ are determined to be\footnote{As expected, it follows from Eq. (\ref{AB2}) that $|c_{+}(x_{1})|^2 - |c_{-}(x_{1})|^2=1$.}:
\begin{equation}
 c_{+}(x_{1}) = \frac{e^{2 i x_1}( 2 x_1^2 -1 + 2 i x_1)}{2 x_1^2},
\qquad
c_{-}(x_{1}) =\frac{1}{2 x_1^2},
\label{AB2}
\end{equation}
where, following the same notation of Eq. (\ref{modf}),  $x_{1} = k\,\tau_{1}$. Recalling Eqs. (\ref{FFGG}) and 
(\ref{FGGF}) and using Eq. (\ref{BogB}) all the power spectra appearing in 
Eq. (\ref{EN6}) can be written in more explicit terms. 

The general formulas are rather lengthy and will not be reported; focussing however on 
a single illustrative combination we shall have expressions of the following kind:
\begin{eqnarray}
{\mathcal P}_{gg}(k,\tau) + k^2 {\mathcal P}_{ff}(k,\tau) &=& \frac{4 \ell_{\mathrm{P}}^2\,\, k^3}{\pi^2\, a^2(\tau)} \biggl\{  (2 |c_{-}(x_1)|^2 + 1) \biggl[ |G_{k}(x,x_{1})|^2 + k^2 |F_{k}(x,x_1)|^2\biggr]
\nonumber\\
&+& c_{-}(x_{1})c_{+}^{\ast}(x_{1}) \biggl[ G^{\ast\, 2}_{k}(x,x_{1}) + k^2 F^{\ast\, 2}_{k}(x,x_{1}) \biggr]
\nonumber\\
&+& c_{-}^{\ast}(x_{1})c_{+}(x_{1}) \biggl[ G^{2}_{k}(x,x_{1}) + k^2 F^{2}_{k}(x,x_{1}) \biggr]\biggr\},
\label{EN7}
\end{eqnarray}
and so on for the other quadratic and quartic combinations appearing in Eqs. (\ref{EN6}) and (\ref{ENN8}). Note that, in Eq. (\ref{EN7}) $G_{k}(\tau) = F_{k}'(\tau) - {\mathcal H} F_{k}(\tau)$,
in analogy with the notation employed for $g_{k}(\tau)$ in Eq. (\ref{T8b}).

Equation (\ref{AB2}) applies when the relevant modes exit the Hubble radius during the inflationary phase and reenter 
in the radiation epoch. Conversely, if the corresponding wavelengths are still larger than the Hubble radius after matter radiation equality, then 
the mode functions of Eq. (\ref{modr3}) should be substituted by:
\begin{eqnarray}
&& F_{k}(\tau) = \frac{1}{\sqrt{2 k}} \biggl( 1 + \frac{i}{x + x_{2} + 2 x_{1}} \biggr) e^{ i (x + x_{2} + 2 x_{1})}, 
\nonumber\\
&& G_{k}(\tau) = \sqrt{\frac{k}{2}} \biggl( i - \frac{3}{(x + x_{2} + 2 x_{1})} - \frac{3i}{( x+ x_{2} + 2 x_{1})^2} \biggr) e^{i (x+ x_{2} + 2 x_{1})},
\label{EN7a}
\end{eqnarray}
where $x_{2} = k \,\tau_{2}$ and $\tau_{2}$ coincides with the equality time; Eq. (\ref{EN7a}) solves Eq. (\ref{MFUN}) for $\tau> \tau_{2}$ (see Eq. (\ref{radmat})). If 
$x_{1} \ll x_{2} \leq 1$, the corresponding wavelength is about to reenter during the matter 
dominated epoch. If  $x_{1} \ll 1$ and $x_{2} \ll 1 $ the relevant wavelengths are larger than the Hubble radius both during the 
radiation-dominated epoch and during the matter dominated phase.
The coefficients $d_{\pm}(x_{1}, x_{2})$ (obeying, $|d_{+}(x_{1}, x_{2})|^2 - |d_{-}(x_{1},x_{2})|^2 = 1$) valid for $\tau >\tau_{2}$ are:
\begin{eqnarray}
d_{+}(x_{1}, x_{2}) &=& \frac{1}{8 (x_{2} + 2 x_{1})^2} [ - 1 - 4 i (x_{2} + 2 x_{1}) + 8 (x_{2} + 2 x_{1})^2 ] e^{-i ( x_{2} + 2 x_{1})},
\nonumber\\
d_{-}(x_{1},x_{2}) &=& - \frac{1}{8 (x_{2} + 2 x_{1})^2}  e^{3 i ( x_{2} + 2 x_{1})},
\label{EN7b}
\end{eqnarray}

We shall now illustrate the expansion required for the explicit evaluation of the 
power spectra of the energy density, of the pressure and of the anisotropic stress. Thanks to Eqs. (\ref{BogB}) and (\ref{AB2})
the following double expansion holds when the wavelengths are larger than the Hubble radius 
at the end of inflation (i.e. $x_{1} \ll 1$) and later on (i.e. $x\ll 1$) :
\begin{eqnarray}
 && \frac{|g_{k}(x,\,x_{1})|^2}{k} = \frac{1}{18\, x_{1}^4}\,\biggl[ x^4 + {\mathcal O}(x^6) \biggr]  
 + \frac{2 }{3 x_{1}^3}\biggl[ x +   {\mathcal O}(x^3)\biggr] + {\mathcal O}(x_{1}^{-2})
\label{gex}\\
&& f_{k}(x,\,x_{1}) g_{k}^{\ast}(x,\,x_{1})  + f_{k}^{\ast}(x,\,x_{1}) g_{k}(x,\,x_{1}) = - \frac{1}{3 x_{1}^4}\biggl[ x^3 + {\mathcal O}(x^{5}) \biggr] 
\nonumber\\
&& - \frac{2}{x_{1}^3} \biggl[ 1 + \frac{2}{3} x^2 + {\mathcal O}(x^4) \biggr] + {\mathcal O}(x_{1}^{-2})
\label{fgex}\\
&& k |f_{k}(x,\,x_{1})|^2 = \frac{1}{2 x_{1}^4} \biggl[ x^2  - \frac{x^{4}}{3} + {\mathcal O}(x^5)\biggr]
+ \frac{2}{x_{1}^3}  \biggl[ x - \frac{2}{3} x^3 + {\mathcal O}(x^{5}) \biggr] + {\mathcal O}(x_{1}^{-2}),
\label{fex}
\end{eqnarray}
where the symbol ${\mathcal O}(...)$ defines the order of magnitude of the subleading terms.
On the basis of Eqs. (\ref{gex}), (\ref{fgex}) and (\ref{fex}) the hierarchy between the different contributions 
can be gauged since, by definition, not only $x_{1} \ll 1$ and  $x \ll 1$, but also $x_{1} \ll x$. Thus, from Eqs.  (\ref{gex})--(\ref{fex}), we shall have 
that 
\begin{equation}
\frac{|g_{k}(x,\,x_{1})|^2}{k} \ll \biggl[f_{k}(x,\,x_{1}) g_{k}^{\ast}(x,\,x_{1})  + f_{k}^{\ast}(x,\,x_{1}) g_{k}(x,\,x_{1}) \biggr] \ll  k \,|f_{k}(x,\,x_{1})|^2.
\label{exm1}
\end{equation}
The results of Eqs. (\ref{gex})--(\ref{fex}) and (\ref{exm1}) can be used to expand more complicated
combinations of mode functions appearing in Eqs. (\ref{EN6}) and (\ref{ENN8}) in powers 
of $k\tau_{1}\ll 1$ and of $|\vec{q}- \vec{k}|\tau_{1} \ll 1$.  For illustration we report 
the leading terms of the expansion for some relevant contributions:
\begin{eqnarray}
|g_{k}(\tau)|^2 \, |g_{|\vec{q} - \vec{k}|}(\tau)|^2 &\simeq& \frac{k |\vec{q} - \vec{k}|}{324} \biggl(\frac{\tau}{\tau_{1}}\biggr)^8,
\nonumber\\
|f_{k}(\tau)|^2 \,  |f_{|\vec{q} - \vec{k}|}(\tau)|^2 &\simeq& \frac{1}{4 k^3 |\vec{q} -  \vec{k}|^3} \frac{\tau^2}{\tau_{1}^4},
\nonumber\\
f_{k}(\tau)g_{k}^{\ast}(\tau) f_{|\vec{q} -\vec{k}|}(\tau)g_{|\vec{q} - \vec{k}|}^{\ast}(\tau) &\simeq& \frac{1}{36 k |\vec{q} - \vec{k}|} \frac{\tau^{3}}{\tau_{1}^4}.
\label{exm2}
\end{eqnarray}
The results of Eq. (\ref{exm2}) have been derived by introducing the four dimensionless variables:
\begin{equation}
 x_{1} = k\tau_{1}\ll 1,\qquad \tilde{x}_{1} = |\vec{q}- \vec{k}|\tau_{1} \ll 1, \qquad x = k\tau \ll 1, \qquad \tilde{x}= |\vec{q}- \vec{k}|\tau \ll 1,
 \label{exm3}
 \end{equation}
and by enforcing the conditions  $x_{1} \ll x \ll 1$ and  $\tilde{x}_{1} \ll \tilde{x} \ll 1$. The results of Eq. (\ref{exm2}) then follow rather easily by  
expressing the leading results of the expansion in terms of the original variables, i.e. ($k$, $|\vec{q} - \vec{k}|$) and 
($\tau$, $\tau_{1}$). 

\subsection{Asymptotic expressions of the power spectra}
Equations (\ref{EN6}) and (\ref{ENN8}) can now be evaluated in the interesting 
physical limits (for instance $x_{1} \ll x \ll 1$ and  $\tilde{x}_{1} \ll \tilde{x} \ll 1$) by using the results of Eqs. (\ref{exm1}) and (\ref{exm2})--(\ref{exm3}). The leading order results for the the expansion are:
\begin{eqnarray}
{\mathcal P}_{\rho_{\mathrm{gw}}}^{(FP)}(q,\tau) &\simeq& \frac{H^8}{512 \, \pi^{5}\, a^4(\tau)} \, |q \tau_{1}|^4\,  {\mathcal F}_{\rho_{\mathrm{gw}}}^{(FP)}(q),
\label{EN8}\\
{\mathcal P}_{\rho_{\mathrm{gw}}}^{(LL)}(q,\tau) &\simeq& \frac{H^8}{512 \, \pi^{5}\, a^4(\tau)} \, |q \tau_{1}|^4\,  {\mathcal F}_{\rho_{\mathrm{gw}}}^{(LL)}(q),
\label{EN11}
\end{eqnarray}
 where ${\mathcal F}_{\rho_{\mathrm{gw}}}^{(FP)}(q)$ and ${\mathcal F}_{\rho_{\mathrm{gw}}}^{(LL)}(q)$
 are defined as:
\begin{eqnarray}
{\mathcal F}_{\rho_{\mathrm{gw}}}^{(FP)}(q) &=& \int d^{3}k \frac{[\vec{k}\cdot\vec{q} - k^2]^2\,[k^2 |\vec{q} - \vec{k}|^2 + 3 (\vec{k}\cdot{q} -k^2)^2][k^2 |\vec{q} - \vec{k}|^2 +  (\vec{k}\cdot{q} -k^2)^2]}{k^7\,q\,|\vec{q} - \vec{k}|^7}.
\nonumber\\
{\mathcal F}_{\rho_{\mathrm{gw}}}^{(LL)}(q) &=& \int d^{3}k \frac{[k^2 |\vec{q} - \vec{k}|^2 + 3 (\vec{k}\cdot\vec{q} -k^2)^2][k^2 |\vec{q} - \vec{k}|^2 +  (\vec{k}\cdot{q} -k^2)^2]}{q\,k^7\,|\vec{q} - \vec{k}|^7}   \biggl[ |\vec{k}\cdot\vec{q} - k^2|^2
\nonumber\\
&+& \frac{16}{9} \biggl(|\vec{q} -\vec{k}|^4 + \frac{16}{9}k^4 \biggr)+ \frac{32}{9} k^2 |\vec{q} -\vec{k}|^2 - \frac{4}{3} \vec{k}\cdot(\vec{q} - \vec{k}) \, ( k^2 + |\vec{q} - \vec{k}|^2) \biggr].
\label{EN12}
\end{eqnarray}
Both ${\mathcal F}_{\rho_{\mathrm{gw}}}^{(FP)}(q)$ and ${\mathcal F}_{\rho_{\mathrm{gw}}}^{(LL)}(q) $
are dimensionless. 
Let us now pause for a moment and remark that 
Eqs. (\ref{EN8}) and (\ref{EN11}) are a direct consequence of  Eqs. (\ref{EN3}) and (\ref{EN4}).  
In spite of the different forms of ${\mathcal F}_{\rho_{\mathrm{gw}}}^{(LL)}(q)$ and ${\mathcal F}_{\rho_{\mathrm{gw}}}^{(FP)}(q)$ their contribution
to the spectrum of the energy density is of comparable magnitude. The integrals appearing in Eq. (\ref{EN12}) can be 
written, in more explicit form, as
\begin{eqnarray}
{\mathcal F}_{\rho_{\mathrm{gw}}}^{(FP)}(q) &=& 2\pi \int_{-1}^{1} d y \int_{k_{\mathrm{min}}}^{k_{\mathrm{max}}} \frac{ k\,d k}{q} \, {\mathcal L}(k,\, q,\,y)
\,( q \,y - k)^2 ,
\label{EN13}\\
{\mathcal F}_{\rho_{\mathrm{gw}}}^{(LL)}(q) &=& \frac{2\pi}{9} \int_{-1}^{1} d y \int_{k_{\mathrm{min}}}^{k_{\mathrm{max}}} \frac{d k}{q\,k} \, {\mathcal L}(k,\, q,\,y)\,{\mathcal M}(k,\, q,\,y),
\label{EN14}
\end{eqnarray}
where the functions ${\mathcal L}(k,\, q,\,y)$ and ${\mathcal M}(k,\, q,\,y)$ are defined as:
\begin{eqnarray}
{\mathcal L}(k,\, q,\,y) &=& \frac{[ 4 k^2 - 8 q \,k\, y + q^2 ( 1 + 3 y^2)] [ 2 k^2 - 4 q\, k\, y + q^2 (1 + y^2)]}{( q^2 + k^2 - 2 q \,k\, y)^{7/2}},
\nonumber\\
{\mathcal M}(k,\, q,\,y) &=&[ 49 k^4 + 16 q^4 + k^2 \, q^2 ( 49 y^2 + 52) - 52 q^3\, k\, y - 62 q\, k^3 \,y].
\label{EN15}
\end{eqnarray}

After performing the explicit integration over $y$  in Eqs. (\ref{EN13}) and (\ref{EN14}), the following results can be obtained: 
\begin{eqnarray}
{\mathcal F}_{\rho_{\mathrm{gw}}}^{(FP)}(q) &=& \frac{8\pi}{105}\int_{k_{\mathrm{min}}}^{k_{\mathrm{max}}} \frac{ d k }{k^6 \, q^2 }\, \biggl[ {\mathcal N}_{+}^{(FP)}(k, q) |k + q| 
+  {\mathcal N}_{-}^{(FP)}(k, q) |k - q| \biggr],
\label{EN16a}\\
{\mathcal F}_{\rho_{\mathrm{gw}}}^{(LL)}(q) &=& \frac{8\pi}{135}\int_{k_{\mathrm{min}}}^{k_{\mathrm{max}}} \frac{ d k }{k^6 \, q^2 }\, \frac{ {\mathcal N}_{-}^{(LL)}(k, q) |k - q| +{\mathcal N}_{+}^{(LL)}(k, q) |k + q| }{ |k + q|\,|k - q|},
\label{EN16}
\end{eqnarray}
where ${\mathcal N}_{\pm}^{(FP)}(k, q)$ and ${\mathcal N}_{\pm}^{(LL)}(k, q)$ are given, respectively, by\footnote{Note that $|k - q|$ and $|k + q|$ do not contain $y$ and differ from $|\vec{k}- \vec{q}| = (k^2 + q^2 - 2 q k y)^{1/2}$.}:
\begin{eqnarray}
{\mathcal N}_{\pm}^{(FP)}(q) &=& \mp 41 k^6 + 169 k^{5} q \mp 169 k^4 q^2 - 146 k^3 q^3 \mp 146 k^2 q^4 + 36 k q^{5} \mp 36 q^{6}
\nonumber\\
{\mathcal N}_{\pm}^{(LL)}(q) &=& \mp 287 k^8 + 1757 k^7 q \mp 1470 k^{6} q^2  - 645 k^{5} q^{3}  \pm 821  k^{4} q^{4} 
\nonumber\\
&+& 278 k^{3} q^{5} \mp 278 k^{2} q^{6} - 36 k q^{7} \mp 36 q^{8}.
\label{EN17}
\end{eqnarray}
Since  Eqs. (\ref{EN16a}) and (\ref{EN16}) are invariant for $k \to k \, \tau_{1}$ and $q\to q \, \tau_{1}$, we can rescale all the integrands through $\tau_{1}$ and expand the obtained result in powers of $|q - k|\tau_{1}$. The result for the derivatives of ${\mathcal F}_{\rho_{\mathrm{gw}}}^{(FP)}(x_{1})$ and ${\mathcal F}_{\rho_{\mathrm{gw}}}^{(LL)}(x_{1})$ with respect to $x_{1}$ will then be:
\begin{eqnarray}
\frac{d {\mathcal F}_{\rho_{\mathrm{gw}}}^{(FP)}}{d x_{1}} &=& \frac{2624 \pi}{105 \, x_{1}} \biggl[ 1 - \frac{2}{x_1} |q - k| \tau_{1}+ \frac{3}{x_{1}^2}  |q - k|^2\tau_{1}^2 + {\mathcal O}( |q - k|^3\tau_{1}^3)\biggr],
\label{EN18}\\
\frac{d {\mathcal F}_{\rho_{\mathrm{gw}}}^{(LL)}}{d x_{1}} &=&\frac{5632 \pi }{135 |q-k|\tau_{1}}\biggl[ 1 +\frac{419 |q-k|\tau_{1}}{88 x_{1}} -\frac{69 |q-k|^2}{x_1^2} +  {\mathcal O}( |q - k|^3\tau_{1}^3)\biggr].
\label{EN19}
\end{eqnarray}
By only keeping the leading terms in the expansions (\ref{EN18}) and (\ref{EN19}) the following pair 
of results shall be obtained:
\begin{eqnarray}
{\mathcal F}_{\rho_{\mathrm{gw}}}^{(FP)}(q) &\simeq& \frac{2624}{105} \pi  \ln{\biggl(\frac{\tau_{2}}{\tau_1}\biggr)},
\label{EN20}\\
{\mathcal F}_{\rho_{\mathrm{gw}}}^{(LL)}(q) &\simeq&\frac{64}{135} \pi  \biggl[331\ln{\biggl(\frac{\tau_{2}}{\tau_1}\biggr)}
-88 \ln{\biggl(\frac{1 - q\tau_{1}}{1 - q\tau_{2}}\biggr)}\biggr],
\label{EN21}
\end{eqnarray}
where the extrema of integration have been taken to be $k_{\mathrm{max}} = 1/\tau_{1}$ and $k_{\mathrm{min}} = 1/\tau_{2}$.
It is natural to require that the largest wavelength (i.e. smallest wavenumber) is the one that will reenter around the equality time
since we are here considering wavelengths that exceed the Hubble radius during the 
radiation dominated phase. In terms of a fiducial set of cosmological parameters, the following expression can be obtained:
\begin{equation}
 \ln{\biggl(\frac{k_{\mathrm{max}}}{k_{\mathrm{min}}}\biggr)} = 58.77 + \ln{{\mathcal B}(r_{T}, {\mathcal A}_{{\mathcal R}},\Omega_{\mathrm{M}0},\, \Omega_{\mathrm{R}0})},
\label{log}
\end{equation}
where\footnote{We recall that $\Omega_{\mathrm{M}0}$ and $\Omega_{\mathrm{R}0}$ are the critical fraction of matter and radiation energy densities; ${\mathcal A}_{{\mathcal R}}$ is the amplitude of the conventional adiabatic mode and $r_{\mathrm{T}}$ is ratio between the inflationary tensor spectrum 
and the spectrum of the adiabatic mode. The typical values correspond to the WMAP 9yr data \cite{WMAP9} (see also \cite{WMAP1}) supplemented by the Bicep2 data \cite{bicep2}.} 
\begin{equation}
{\mathcal B}(r_{T}, {\mathcal A}_{{\mathcal R}},\Omega_{\mathrm{M}0},\, \Omega_{\mathrm{R}0}) = \biggl(\frac{r_{T}}{0.2}\biggr)^{1/4} \biggl(\frac{{\mathcal A}_{\mathcal R}}{2.41 \times 10^{-9}}\biggr)^{1/4}\biggl(\frac{h_{0}^2 \Omega_{\mathrm{R}0}}{4.15 \times 10^{-5}}\biggr)^{3/4}\, \biggl(\frac{h_{0}^2 \Omega_{\mathrm{M}0}}{0.1364}\biggr)^{-1};
\end{equation} 
with the above choice of fiducial values we have that $\log{(\tau_{2}/\tau_{1})} \simeq {\mathcal O}(60)$ in Eqs. (\ref{EN20})--(\ref{EN21}).

The results reported so far demonstrate explicitly that the Landau-Lifshitz and in the Ford-Parker parametrizations are  fully compatible  insofar as they lead to equivalent results.
Barring for potential differences in the numerical prefactor we are led to the following expressions 
\begin{equation}
{\mathcal P}^{(FP)}_{\rho_{\mathrm{gw}}}(q,\tau) = {\mathcal C}^{(FP)}_{\rho_{\mathrm{gw}}} \frac{H^8}{{a^4(\tau)}} q^4 \,\tau_{1}^4, 
\qquad {\mathcal P}^{(LL)}_{\rho_{\mathrm{gw}}}(q,\tau) = {\mathcal C}^{(LL)}_{\rho_{\mathrm{gw}}} \frac{H^8}{{a^4(\tau)}} q^4 \,\tau_{1}^4, 
\label{EN21a}
\end{equation}
with ${\mathcal C}^{(FP)}_{\rho_{\mathrm{gw}}} \neq {\mathcal C}^{(LL)}_{\rho_{\mathrm{gw}}}$. This result 
is more than sufficient for the explicit solution of Eqs. (\ref{GS8}) and (\ref{GS10}). In what follows we shall therefore 
remove the superscripts specifying the parametrization of the energy-momentum 
pseudo-tensor and acknowledge that the net result of the whole calculation can be expressed as follows 
\begin{eqnarray}
{\mathcal P}_{\rho_{\mathrm{gw}}}(q,\tau) &=& {\mathcal C}_{\rho_{\mathrm{gw}}} H^8 \,\biggl[\frac{a_{1}}{a(\tau)}\biggr]^4 \,q^4 \,\tau_{1}^4, 
\nonumber\\
{\mathcal P}_{p_{\mathrm{gw}}}(q,\tau) &=& {\mathcal C}_{p_{\mathrm{gw}}} H^8 \,\biggl[\frac{a_{1}}{a(\tau)}\biggr]^4 q^4 \,\tau_{1}^4, 
\nonumber\\
{\mathcal P}_{\Pi_{\mathrm{gw}}}(q,\tau) &=& {\mathcal C}_{\Pi_{\mathrm{gw}}}  H^8 \,\biggl[\frac{a_{1}}{a(\tau)}\biggr]^4 q^4 \,\tau_{1}^4.
\label{EXP1}
\end{eqnarray}
The constants ${\mathcal C}_{\rho_{\mathrm{gw}}}$, ${\mathcal C}_{p_{\mathrm{gw}}}$ and ${\mathcal C}_{\Pi_{\mathrm{gw}}}$
do depend on the parametrization but they are of the same order.  
Equation (\ref{EXP1}) contains also the results for the pressure and the anisotropic stress that 
are derived by following exactly the same strategy detailed above 
for the case of the energy density. To avoid lengthy digressions the corresponding results for the pressure and the 
anisotropic stress have been collected in the appendix (see, in particular, appendix \ref{appA} for the pressure and appendix \ref{appB}
for the anisotropic stress). When $\tau > \tau_{2}$, Eq. (\ref{EXP1}) keeps in practice the same form
(by replacing $\tau_{1}$ with $\tau_{2}$ and $a_{1}$ with $a_{2}$); the overall constants appearing in the new formulas will then be slightly different and the result will now hold in the limit 
\begin{equation}
\tau > \tau_{2}>| \tau_{1}|,   \qquad |q \tau| \ll 1,\qquad |q \tau_{2}| \ll 1, \qquad |\vec{q} - \vec{k}|\tau_{2} \ll 1.
\label{EXP3a}
\end{equation}

Let us finally consider the following pair of evolution equations obtained, respectively, in the Ford-Parker and in the Landau-Lifshitz parametrizations:
\begin{eqnarray}
\partial_{\tau} \delta \rho_{\mathrm{gw}}^{(FP)} + 3 {\mathcal H} [  \delta \rho_{\mathrm{gw}}^{(FP)} + \delta p_{\mathrm{gw}}^{(FP)}]  - \vec{\nabla}\cdot\vec{{\mathcal J}} =0,
\label{CCFP}\\
\partial_{\tau} \delta \rho_{\mathrm{gw}}^{(LL)} + 3 {\mathcal H} [\delta \rho_{\mathrm{gw}}^{(LL)} + \delta P_{\mathrm{gw}}^{(LL)}]  - \vec{\nabla}\cdot\vec{{\mathcal J}} =0.
\label{CCLL}
\end{eqnarray}
Equation (\ref{CCFP}) comes from Eq. (\ref{FF1}) by requiring $\overline{\nabla}_{\mu} {\mathcal W}^{\mu}_{\nu}=0$, where $\overline{\nabla}_{\mu}$ denotes the covariant derivative defined in terms of the conformally flat metric $\overline{g}_{\mu\nu}$. Equation (\ref{CCLL}) is instead derived
from the second-order (tensor) fluctuation of the Bianchi identity, i.e. $\delta_{\rm t}^{(2)} ( \nabla_{\mu} {\cal G}^{\mu}_{\nu}) =0$.
The Bianchi identity itself (i.e. $\nabla_{\mu} {\cal G}_{\nu}^{\mu}=0$) is valid to all orders and it will be satisfied, 
in particular, to second-order.  As it can be directly checked, Eqs. (\ref{CCFP}) and (\ref{CCLL}) are exactly equivalent 
to Eq. (\ref{MFUN}) once the corresponding definitions of the various fluctuations are taken into account.

The specific form of  $\delta P_{\mathrm{gw}}^{(LL)}$ is given in appendix \ref{appA} (see Eq. (\ref{P8})) and, unlike 
$\delta p_{\mathrm{gw}}^{(FP)}$, it contains an effective viscous contribution. Some authors \cite{murphy,bel}
suggested in the past that the relic gravitons may lead to an effective bulk viscosity in the early Universe. This is not exactly the viewpoint adopted here where the emphasis 
is on those results that are independent on the parametrization of the energy-momentum pseudo-tensor. 
In the long wavelength limit, the last terms in Eqs. (\ref{CCFP})--(\ref{CCLL}) are subleading in comparison with the other contributions of the corresponding 
equations. This occurrence is compatible with the form of the power spectra of Eqs. (\ref{EXP1}) and 
and it implies, in the long wavelength limit (and in spite of the parametrization) that $\delta p_{\mathrm{gw}} \simeq - \delta \rho_{\mathrm{gw}}/3$ and that $\delta \rho_{\mathrm{gw}} \simeq a^{-2}$. 

 \subsection{The induced curvature perturbations}

Equations (\ref{EXP1}) and (\ref{CCFP})--(\ref{CCLL}) 
are essential for the explicit evaluation of the integrals appearing in  Eq. (\ref{GS10}).
The first line of Eq. (\ref{GS10}) contains two factors: the adiabatic piece (which is left untouched by the present 
discussion) and a further contribution (depending on the first time derivative of ${\mathcal R}$ at $\tau_{1}$). This term decreases in time 
and it can be soon neglected for $\tau > \tau_{1}$. The latter conclusion follows by observing that $z_{\mathrm{t}}(\tau)$ is proportional
to $a(\tau)$: during radiation (i.e. ${\mathcal H}' = - {\mathcal H}^2$) 
$z_{\mathrm{t}}^2(\tau)= 12 a^2(\tau)/\ell_{\mathrm{P}}^2$. This result implies that 
\begin{equation}
z_{\mathrm{t}}^2(\tau_{1})\int_{\tau_{1}}^{\tau} \frac{d \tau'}{z_{\mathrm{t}}^2(\tau')} = a^2(\tau_{1}) {\mathcal H}_{1} \biggl( 1 - \frac{\tau_{1}}{\tau}\biggr).
\label{EXP2}
\end{equation}
Moreover, in the purely adiabatic case, ${\mathcal R}'(k,\tau_{1}) =0$ and therefore the prefactor multiplying 
Eq. (\ref{EXP2}) only depends on $\Sigma_{\mathcal R}(k,\tau_{1})$. 

Having discussed the first line of Eq. (\ref{GS10}) we are left with the remaining terms that can be written as:
\begin{eqnarray}
{\mathcal I}_{1}(\vec{q}, \tau) &=&   \int_{\tau_{1}}^{\tau} \Sigma_{{\mathcal R}}(\vec{q},\tau') \, \, d\tau',
\label{EXP3}\\
{\mathcal I}_{2}(\vec{q}, \tau) &=& 3 \int_{\tau_{1}}^{\tau} \frac{ d \tau'}{z_{\mathrm{t}}^2(\tau')} \int_{\tau_{1}}^{\tau'} a^4(\tau^{\prime\prime}) \, \Pi_{\mathrm{gw}}(\vec{q},\tau^{\prime\prime}).
\label{EXP4}
\end{eqnarray}
Owing to the specific form of the power spectra of Eq. (\ref{EXP1}) and thanks to Eqs.  (\ref{CCFP})--(\ref{CCLL}) we can also write the Fourier space components of the various fluctuations in the following factorized form:
\begin{equation}
\delta \rho_{\mathrm{gw}}(\vec{q},\tau) = \frac{\delta \rho_{\mathrm{gw}}(\vec{q}) }{a^2(\tau)},\qquad \delta p_{\mathrm{gw}}(\vec{q},\tau) = \frac{\delta p_{\mathrm{gw}}(\vec{q}) }{a^2(\tau)},\qquad \Pi_{\mathrm{gw}}(\vec{q},\tau) = \frac{\Pi_{\mathrm{gw}}(\vec{q})}{a^2(\tau)}.
\label{EXP5}
\end{equation}
Thanks to Eq. (\ref{EXP5}) the integrals of Eqs. (\ref{EXP3})--(\ref{EXP4}) can be performed in explicit terms and the result is:
\begin{equation}
{\mathcal I}_{1}(\vec{q},a) = \frac{{\mathcal E}_{1}(\vec{q})}{H^2 \overline{M}_{\mathrm{P}}^2 } {\mathcal D}_{1}(w, a), \qquad {\mathcal I}_{2}(\vec{q},a)= \frac{{\mathcal E}_{2}(\vec{q})}{H^2 \overline{M}_{\mathrm{P}}^2 }{\mathcal D}_{2}(w, a),
\label{EXP5a}
\end{equation}
where $\overline{M}_{\mathrm{P}} = \ell_{\mathrm{P}}^{-1}$ and\footnote{In principle the interesting cases are $w=1/3$ and $w=0$. However,
the analysis reported in the preceding section can be generalized to the case of a generic post-inflationary phase with the result 
that Eq. (\ref{EXP1})  and (\ref{CCFP})--(\ref{CCLL})  are formally valid.}
\begin{eqnarray}
&& {\mathcal E}_{1}(\vec{q}) = c_{\mathrm{st}}^2 \delta \rho_{\mathrm{gw}}(\vec{q}) - \delta p_{\mathrm{gw}}(\vec{q}) + \Pi_{\mathrm{gw}}(\vec{q}), \qquad {\mathcal E}_{2}(\vec{q}) = \Pi_{\mathrm{gw}}(\vec{q}),
\nonumber\\
&& {\mathcal D}_{1}(w,a) = \frac{1}{3 (3 w +1) ( w + 1) }\biggl[ \biggl(\frac{a}{a_{1}}\biggr)^{3 w +1} -1 \biggr],
\nonumber\\
&& {\mathcal D}_{2}(w,a) =  \frac{w ( 3 w+1)}{6(w -1)(w+1)} \biggl[1 + \frac{3 ( w-1)}{(3 w + 5)}
\biggl(\frac{a}{a_{1}}\biggr)^{3 w +1} - \frac{2(3 w + 1)}{(3 w +5)} \biggl(\frac{a}{a_{1}}\biggr)^{3(w -1)/2} \biggr].
\nonumber
\end{eqnarray}
The result for $ {\mathcal D}_{2}(w,a)$ is correct provided $w\neq 1$; in the case $w=1$ (stiff post-inflationary phase) the results 
are physically equivalent\footnote{In the case of a stiff post-inflationary phase ${\mathcal D}_{1}(1, a) = [(a/a_{1})^2-1]/24$ and 
${\mathcal D}_{2}(1, a) = [(a/a_1)^4 -1 - 2 \ln{(a/a_{1})}]/4$. The analysis of the correlation functions can be conducted also in this 
case with results that are quantitatively compatible with the ones of the radiation case. This analysis will not be reported in detail since 
it is not central to the theme of this paper.}. 

Using the results of Eqs. (\ref{EXP3}), (\ref{EXP4}) and (\ref{EXP5a}) the power spectrum of curvature perturbations is given by
\begin{eqnarray}
{\mathcal P}_{{\mathcal R}}(q,\tau) &=& {\mathcal A}_{{\mathcal R}} + \biggl(\frac{H}{\overline{M}_{\mathrm{P}}}\biggr)^4\, q^4 \tau_{1}^4 \,\biggl[ 
{\mathcal C}_{1} {\mathcal D}_{1}(w,a) + {\mathcal C}_{2} {\mathcal D}_{2}(w,a)\biggr]^2,
\label{EXP9}\\
{\mathcal C}_{1}&=& c_{\mathrm{st}}^2 \sqrt{{\mathcal C}_{\rho_{\mathrm{gw}}}} - \sqrt{{\mathcal C}_{p_{\mathrm{gw}}}} + \sqrt{{\mathcal C}_{\Pi_{\mathrm{gw}}}},\qquad 
{\mathcal C}_{2} = \sqrt{{\mathcal C}_{\Pi_{\mathrm{gw}}}}.
\label{EXP11}
\end{eqnarray}
Consider now two specific (but important) cases, namely $w=1/3$ (radiation-dominated case) and $w=0$ (matter dominated case\footnote{Notice that if $ w\to 0$ ${\mathcal D}_{2} \to 0$ but ${\mathcal D}_{1} \neq 0$.}).
In the radiation case, 
recalling that $H/\mathrm{M}_{\mathrm{P}} = \pi \, \sqrt{r_{T} {\mathcal A}_{{\mathcal R}}/2}$, we have:
\begin{equation}
{\mathcal P}_{{\mathcal R}}(q,\tau) \simeq {\mathcal A}_{{\mathcal R}} + \frac{r_{T}^2}{576}\, {\mathcal A}_{{\mathcal R}}^2 q^4 \tau_{1}^4 ( 3 {\mathcal C}_{1} + {\mathcal C}_{2})^2\biggl(\frac{a}{a_{1}}\biggr)^{4}.
\label{EXP12}
\end{equation}
The wavelength of the fluctuation becomes comparable with the Hubble radius 
at the reentry time namely when $q \tau_{\mathrm{re}} =1$. Let us suppose, as implied by Eq. (\ref{EXP12}) that the given mode reenters  during the radiation-dominated phase:
\begin{equation}
{\mathcal P}_{{\mathcal R}}(q,\tau_{\mathrm{re}}) \simeq {\mathcal A}_{{\mathcal R}} + \frac{r_{T}^2}{576}\, {\mathcal A}_{{\mathcal R}}^2 ( 3 {\mathcal C}_{1} + {\mathcal C}_{2})^2 q^4 \tau_{1}^4 \biggl(\frac{\tau_{\mathrm{re}}}{\tau_{1}}\biggr)^{4},
\label{EXP13}
\end{equation}
since $q \tau_{1} (a_{\mathrm{re}}/a_{1}) = q \tau_{\mathrm{re}}$ (and, by definition, $q \tau_{\mathrm{re}} =1$) Eq. (\ref{EXP13})
also implies that 
\begin{equation}
{\mathcal P}_{{\mathcal R}}(q,\tau_{\mathrm{re}}) \simeq {\mathcal A}_{{\mathcal R}} + \frac{9\,{\mathcal C}_{\rho_{\mathrm{gw}}}}{144}\, r_{T}^2\,{\mathcal A}_{{\mathcal R}}^2.
\label{EXP14}
\end{equation}
In Eq. (\ref{EXP14}) we used the following chain of equalities:
\begin{equation}
\frac{1}{576} ( 3 {\mathcal C}_{1} + {\mathcal C}_{2})^2 \simeq \frac{1}{144} (\sqrt{{\mathcal C}_{\rho_{\mathrm{gw}}}} + 2 \sqrt{{\mathcal C}_{\Pi_{\mathrm{gw}}}})^2 \simeq \frac{9\, {\mathcal C}_{\rho_{\mathrm{gw}}}}{144} .
\label{EXP15}
\end{equation}
The last equality at the right 
hand side of Eq. (\ref{EXP15}) 
follows by assuming  that ${\mathcal C}_{\Pi_{\mathrm{gw}}}$ and ${\mathcal C}_{\rho_{\mathrm{gw}}}$ are of the same order.
On the basis of the estimates previously discussed within the different parametrizations, the whole numerical factor multiplying 
$ r_{T}^2\,{\mathcal A}_{{\mathcal R}}^2$ in the second term at the right hand side of Eq. (\ref{EXP14})  can be estimated between 
$2\times 10^{2}$ and $5 \times 10^{2}$. Finally, if the given mode reenters during the the matter-dominated phase we have 
\begin{equation}
{\mathcal P}_{{\mathcal R}}(q,\tau_{\mathrm{re}}) \simeq {\mathcal A}_{{\mathcal R}} + \frac{9\,\overline{{\mathcal C}}_{\rho_{\mathrm{gw}}} }{144}\, r_{T}^2\,{\mathcal A}_{{\mathcal R}}^2   q^4 \tau_{1}^4 \biggl(\frac{a_{\mathrm{eq}}}{a_{1}}\biggr)^{4}\,\biggl(\frac{a_{\mathrm{re}}}{a_{\mathrm{eq}}}\biggr)^{2},
\label{EXP14a}
\end{equation}
which gives back the result of Eq. (\ref{EXP13}) since, during matter-dominated phase, $(a_{\mathrm{re}}/a_{\mathrm{eq}})^2 \simeq 
(\tau_{\mathrm{re}}/\tau_{\mathrm{eq}})^4$. In Eq. (\ref{EXP14a}) we defined $\overline{{\mathcal C}}_{\rho_{\mathrm{gw}}}$ which is numerically different from ${\mathcal C}_{\rho_{\mathrm{gw}}}$ but it has the same origin and it is of the same order.

\newpage

\renewcommand{\theequation}{5.\arabic{equation}}
\setcounter{equation}{0}
\section{Concluding remarks}
\label{sec5}
The fluctuations of the energy density of the long and short wavelength gravitons are 
strongly suppressed relative to the background energy density which is 
roughly constant during the inflationary phase. The same statement holds for the fluctuations of the pressure and for the anisotropic 
stress. Conversely the same fluctuations induce curvature perturbations that grow during the post-inflationary epoch. 

Barring for some evolution in the number of relativistic species of the plasma, after the end of inflation 
the short wavelength gravitons produce scalar inhomogeneities that decrease roughly at the same rate of the radiation 
background: if they were small at the end of inflation they will remain small. 
The fluctuations of the energy density, pressure and anisotropic stress of the long wavelength 
gravitons decrease in time at a rate that it is slower than the one  
of the radiation and matter energy densities. In this situation the curvature 
perturbations inherit a a supplementary contribution that depends on the ratio between the fluctuations of the energy momentum pseudo-tensor of the 
relic gravitons and the background energy density. 

The corrections to the power spectrum of the curvature perturbations can then be 
computed and they mildly depend on the parametrizations 
of the energy momentum pseudo-tensor. Since the definition 
of the momentum and energy of the gravitational field itself is formally not unique, a more pragmatic approach has been adopted. The same analysis 
has been performed using completely different assignments  of the energy-momentum 
pseudo-tensor. A posteriori, the compatibility of the obtained results fully justifies our strategy. It turns out that the growth of the 
curvature perturbations induced by this effect is approximately compensated by the corresponding spectral slope.

All in all, we argue that the fluctuations of the long-wavelength gravitons can be treated in the fluid approximation 
but different pseudo-tensors suggest slightly different properties of the effective fluid.
For instance the classic Landau-Lifshitz parametrization (appropriately generalized to the case of conformally flat backgrounds) suggests that 
the gravitons contribute to an effective bulk viscosity of the corresponding fluid. If we choose instead
to parametrize the energy-momentum pseudo-tensor in terms of two minimally coupled scalar fields 
(one for each polarization of  the graviton) the effective fluid is inviscid. While the optimal or parametrization 
of the energy-momentum pseudo-tensor of the relic gravitons is not at issue, the present findings show, more 
modestly, that different strategies lead, in practice, to consistent physical results. 

\newpage 
\begin{appendix}

\renewcommand{\theequation}{A.\arabic{equation}}
\setcounter{equation}{0}
\section{Power spectrum of the energy density}
\label{appAA}
The power spectrum of the energy density in the Landau-Lifshitz 
parametrization has been discussed in the bulk of the paper but we report here,
for the sake of precision, the explicit expressions of  ${\mathcal A}_{1}(\vec{k},\vec{q},\tau)$, ${\mathcal A}_{2}(\vec{k},\vec{q},\tau)$ and 
${\mathcal A}_{3}(\vec{k},\vec{q},\tau)$ appearing in Eq. (\ref{ENN8}):
\begin{eqnarray}
 {\mathcal A}_{1}(\vec{k},\vec{q},\tau) &=& {\mathcal P}_{ff}(k,\tau)  {\mathcal P}_{gg}(|\vec{q} - \vec{k}|,\tau) + {\mathcal P}_{ff}(|\vec{q} - \vec{k}|,\tau){\mathcal P}_{gg}(k,\tau)
\nonumber\\
&+& {\mathcal P}_{fg}(k,\tau) {\mathcal P}_{gf}(|\vec{q} - \vec{k}|,\tau) + {\mathcal P}_{fg}(|\vec{q} - \vec{k}|,\tau) {\mathcal P}_{gf}(k,\tau),
\label{ENN9}\\
{\mathcal A}_{2}(\vec{k},\vec{q},\tau) &=&  {\mathcal P}_{fg}(k,\tau) {\mathcal P}_{gg}(|\vec{q} - \vec{k}|,\tau) + 
 {\mathcal P}_{fg}(|\vec{q} - \vec{k}|,\tau)  {\mathcal P}_{gg}(k,\tau) 
 \nonumber\\
&+& {\mathcal P}_{gf}(k,\tau)  {\mathcal P}_{gg}(|\vec{q} - \vec{k}|,\tau) 
 +  {\mathcal P}_{gf}(|\vec{q} - \vec{k}|,\tau)   {\mathcal P}_{gg}(k,\tau) 
 \nonumber\\
 &+& [\vec{k}\cdot(\vec{q} - \vec{k})] \, \biggl[   {\mathcal P}_{ff}(k,\tau) \biggl( {\mathcal P}_{gf}(|\vec{q} - \vec{k}|,\tau) +  {\mathcal P}_{fg}(|\vec{q} - \vec{k}|,\tau)\biggr)
 \nonumber\\
&+&  {\mathcal P}_{ff}(|\vec{q}- \vec{k}|,\tau)\biggl( {\mathcal P}_{gf}(k,\tau) +  {\mathcal P}_{fg}(k,\tau)\biggr)\biggr],
 \label{ENN10}\\
  {\mathcal A}_{3}(\vec{k},\vec{q},\tau) &=& {\mathcal P}_{gg}(k,\tau)  {\mathcal P}_{gg}(|\vec{q} - \vec{k}|,\tau) + 
 [ \vec{k}\cdot(\vec{q} - \vec{k})]^2 \,  {\mathcal P}_{ff}(k,\tau)  {\mathcal P}_{ff}(|\vec{q} - \vec{k}|,\tau)
\nonumber\\
 &+& \vec{k}\cdot(\vec{q} - \vec{k})\, \biggl({\mathcal P}_{fg}(k,\tau){\mathcal P}_{fg}(|\vec{q} - \vec{k}|,\tau) + {\mathcal P}_{gf}(k,\tau){\mathcal P}_{gf}(|\vec{q} - \vec{k}|,\tau)
\biggr).
\label{ENN11}
\end{eqnarray}
\renewcommand{\theequation}{B.\arabic{equation}}
\setcounter{equation}{0}
\section{Power spectrum of the pressure}
\label{appA}
The same strategy adopted for the calculation of the fluctuations of the energy density can be 
pursued to compute the fluctuations of the pressure. 
The discussion on the differences between the two parametrizations shall not be repeated here since 
it mirrors exactly the results mentioned in the case 
of the energy density. The fluctuations 
of the pressure and the corresponding two-point function are defined exactly as in the case of the energy density:
\begin{eqnarray}
&&  \delta p_{\mathrm{gw}}(\vec{x},\tau) = \frac{1}{(2\pi)^{3/2}} \int d^{3} k \, e^{- i \vec{k} \cdot \vec{x}}\, \delta p_{\mathrm{gw}}(\vec{k},\tau),
\label{P1}\\
&& \langle \delta p_{\mathrm{gw}}(\vec{p},\tau) \delta p_{\mathrm{gw}}(\vec{q},\tau) \rangle = \frac{2\pi^2}{q^3}\, 
\delta^{(3)}(\vec{q} + \vec{p}) \, {\mathcal P}_{p_{\mathrm{gw}}}(q,\tau).
\label{P2}
\end{eqnarray}
The fluctuations of the pressure in the two parametrizations are, respectively,
\begin{eqnarray}
\delta p^{(FP)}_{\mathrm{gw}}(\vec{q},\tau) &=& \frac{1}{8 \ell_{\mathrm{P}}^2 a^2 } \int \frac{d^{3} k}{(2\pi)^{3/2}} \biggl[ \hat{H}_{\ell m}(\vec{k},\tau) 
 \hat{H}_{\ell m}(\vec{q} -\vec{k},\tau)
 \nonumber\\
 &-& \frac{(\vec{k} - \vec{q})\cdot\vec{k}}{3}  \,\hat{h}_{\ell m}(\vec{k},\tau) 
 \hat{h}_{\ell m}(\vec{q} -\vec{k},\tau) \biggr],
\label{P3}\\
 \delta p^{(LL)}_{\mathrm{gw}}(\vec{q},\tau) &=& - \frac{1}{24 \ell_{\mathrm{P}}^2 a^2 } \int \frac{d^{3} k}{(2\pi)^{3/2}}\biggl[  5 \hat{H}_{\ell m}(\vec{k},\tau) 
 \hat{H}_{\ell m}(\vec{q} -\vec{k},\tau) 
 \nonumber\\
 &-& 7 [(\vec{k} - \vec{q})\cdot\vec{k}]\, \hat{h}_{\ell m}(\vec{k},\tau)\hat{h}_{\ell m}(\vec{q} -\vec{k},\tau)\biggr].
 \label{P4}
 \end{eqnarray}
We first analyze the correlation of the pressure in the Ford-Parker parametrization. In this case the approximate result for the power 
spectrum of the pressure is:
\begin{eqnarray}
{\mathcal P}^{(FP)}_{p_{\mathrm{gw}}}(q,\tau) = \frac{q^3}{1152 \pi^5 \ell_{\mathrm{P}}^4 a^4} \int \frac{d^{3} k}{k^3 \,|\vec{q} - \vec{k}|^3}\, [\vec{k}\cdot(\vec{k} - \vec{q})]^2\, Q(\vec{k},\vec{q})\,  P_{ff}(k,\tau)\, P_{ff}(|\vec{q} - \vec{k}|,\tau).
\label{P5}
\end{eqnarray}
The result obtained for the pressure can be expressed by saying that the power spectra of the energy density 
and of the pressure are related as:
\begin{equation}
{\mathcal P}^{(FP)}_{p_{\mathrm{gw}}}(q,\tau) \simeq \frac{{\mathcal P}^{(FP)}_{\rho_{\mathrm{gw}}}(q,\tau)}{9}.
\label{P6}
\end{equation}
This result can be understood on the basis of Eq. (\ref{CCFP}):
\begin{equation}
\partial_{\tau} \delta\rho_{\mathrm{gw}}^{(FP)} + 3 {\mathcal H} [\delta\rho_{\mathrm{gw}}^{(FP)} + \delta p_{\mathrm{gw}}^{(FP)}] -
\vec{\nabla}\cdot \vec{{\mathcal J}} =0.
\label{P7}
\end{equation}
The third term in Eq. (\ref{P7}) is subleading in comparison with the other two when the relevant wavelengths exceed the Hubble radius 
at the corresponding epoch. Now, the evolution of the power spectrum of the energy density 
deduced in section \ref{sec4} implies that $\delta\rho_{\mathrm{gw}}^{(FP)}(\vec{q},\tau)$ scales as $a^{-2}$ which 
is compatible with Eq. (\ref{P7}) provided the effective barotropic index when the relevant wavelengths exceed the 
Hubble radius is ${\mathcal O}(-1/3)$. This result is in turn compatible with Eq. (\ref{P6}).

As we saw in section \ref{sec4} the different parametrizations are physically equivalent and differ, at the level of the power spectrum, just 
by the specific value of the overall constant. If this is true we expect that the same result of Eq. (\ref{P7}) can also 
be obtained in the framework of the Landau-Lifshitz parametrization where 
 the evolution of the energy density and pressure fluctuation comes by perturbing to second 
order the Bianchi identity that must hold at any order in the perturbative expansion. The result 
of this procedure implies that the equation obeyed is\footnote{It should be clear that $\delta P_{\mathrm{gw}}^{(LL)}$ 
differs from ${\mathcal P}_{\mathrm{gw}}^{(LL)}$ and $\overline{{\mathcal P}}_{\mathrm{gw}}^{(LL)}$. While  $\delta P_{\mathrm{gw}}^{(LL)}$  denotes the pressure 
fluctuation in the Landau-Lifshitz parametrization (and in the presence of viscous contribution, see Eq. (\ref{P8})), ${\mathcal P}_{p_\mathrm{gw}}^{(LL)}$ and $\overline{{\mathcal P}}_{P_\mathrm{gw}}^{(LL)}$ are the power spectra defined, respectively,  from $\delta p_{\mathrm{gw}}^{(LL)}$ and $\delta P_{\mathrm{gw}}^{(LL)}$. With this caveat any potential clash of notation is avoided.}
\begin{equation}
\partial_{\tau} \delta\rho_{\mathrm{gw}}^{(LL)} + 3 {\mathcal H} [\delta\rho_{\mathrm{gw}}^{(LL)} + \delta P_{\mathrm{gw}}^{(LL)}]
-\vec{\nabla}\cdot \vec{{\mathcal J}} =0.
\label{PP7}
\end{equation}
where $\delta P_{\mathrm{gw}}^{(LL)}$ does not coincide with  $\delta p_{\mathrm{gw}}^{(LL)}$ given in Eq. (\ref{P4}) but it is defined as
\begin{equation}
\delta P_{\mathrm{gw}}^{(LL)} = \delta p_{\mathrm{gw}}^{(LL)} + \frac{{\mathcal H}^2 - {\mathcal H}'}{ 3 {\mathcal H} a^2 \ell_{\mathrm{P}}^2} H_{\ell m} 
h_{\ell m}. 
\label{P8}
\end{equation}
We can therefore compute the power spectra of $\delta p_{\mathrm{gw}}^{(LL)}$ and of $\delta P_{\mathrm{gw}}^{(LL)}$. Defining 
${\mathcal P}^{(LL)}_{p_{\mathrm{gw}}}(q,\tau)$ as the power spectrum of $\delta p_{\mathrm{gw}}^{(LL)}$  we have: 
\begin{equation}
{\mathcal P}^{(LL)}_{p_{\mathrm{gw}}}(q,\tau) \simeq \frac{49}{25} {\mathcal P}^{(LL)}_{\rho_{\mathrm{gw}}}(q,\tau).
\label{P9}
\end{equation}
This result can be obtained by using exactly the same techniques described in the previous section. 
However, defining $\overline{{\mathcal P}}^{(LL)}_{P_{\mathrm{gw}}}(q,\tau)$ as the power spectrum 
of $\delta P_{\mathrm{gw}}^{(LL)} $ we have:
\begin{equation}
\overline{{\mathcal P}}^{(LL)}_{P_{\mathrm{gw}}}(q,\tau) \simeq \frac{1}{3}{\mathcal P}^{(LL)}_{\rho_{\mathrm{gw}}}(q,\tau).
\label{P10}
\end{equation}
The dictionary between the two parametrizations of the energy-momentum pseudo-tensor implies therefore 
that $\delta p_{\mathrm{gw}}^{(FP)}$ is directly comparable with $\delta P_{\mathrm{gw}}^{(LL)}$  (rather than 
with $\delta p_{\mathrm{gw}}^{(LL)}$).
\renewcommand{\theequation}{C.\arabic{equation}}
\setcounter{equation}{0}
\section{Power spectrum of the anisotropic stress}
\label{appB}

In spite of the specific parametrization of the anisotropic stress we shall have that 
\begin{equation}
\Pi_{\mathrm{gw}}(\vec{q},\tau) = \frac{q_{i} q_{j}}{q^2} \Pi_{ij}(\vec{q},\tau). 
\end{equation}
The fluctuations of the energy density in the Ford-Parker and Landau-Lifshitz 
parametrizations are, respectively:
\begin{eqnarray}
\Pi^{(FP)}_{\mathrm{gw}}(\vec{q},\tau) &=&\frac{1}{12 \ell_{\mathrm{P}}^2 a^2}\int \frac{d^{3}k}{(2\pi)^{3/2}} \biggl[ 2 (\vec{q}\cdot\vec{k}) - 
3 \frac{(\vec{q}\cdot\vec{k})^2}{q^2} + k^2 \biggr]\, \hat{h}_{\ell m}(\vec{k},\tau)\,\hat{h}_{\ell m}(\vec{q} - \vec{k},\tau),
\nonumber\\
\Pi^{(LL)}_{\mathrm{gw}}(\vec{q},\tau) &=&\frac{1}{12 \ell_{\mathrm{P}}^2 a^2}\int \frac{d^{3}k}{(2\pi)^{3/2}} \biggl\{ 2   \hat{H}_{\ell m}(\vec{k},\tau) \hat{H}_{\ell m}(\vec{q} - \vec{k},\tau)
\nonumber\\
&+&\biggl[ \vec{k}\cdot(\vec{q} - \vec{k}) + 3 (\vec{k}\cdot\vec{q}) - 3 \frac{(\vec{k}\cdot\vec{q})^2}{q^2} \biggr]  \hat{h}_{\ell m}(\vec{k},\tau) \hat{h}_{\ell m}(\vec{q} - \vec{k},\tau) 
\nonumber\\
&-& 6 \frac{q^{i} q^{j}}{q^2} [ \vec{k}\cdot(\vec{q} - \vec{k}) \hat{h}_{m i}(\vec{k},\tau) \hat{h}_{m j}(\vec{q} - \vec{k},\tau) - \hat{h}_{m i}(\vec{k},\tau) 
\hat{h}_{m j}(\vec{q} - \vec{k},\tau)] \biggr\}.
\label{LLan}
\end{eqnarray}
The power spectrum of the anisotropic stress will be only compute in the Ford-Parker parametrization. As in the case 
of the energy density and of the pressure we have that 
\begin{eqnarray}
{\mathcal P}_{\Pi}(q,\tau) &=& \frac{ |q\,\tau_{1}|^4 }{1152 \, \pi^{5}} \frac{H^8}{a^4} \, {\mathcal F}_{\Pi}^{(FP)}(q),
\label{ANIS1}\\
{\mathcal F}_{\Pi}^{(FP)}(q) &=& \int d^{3} k \frac{[ 2 (\vec{q} \cdot\vec{k})\,q^2 - 3(\vec{q}\cdot\vec{k})^2 + k^2 q^2]^2 }{q^{5} \, k^{7}\, |\vec{q} -\vec{k}|^7}
\nonumber\\
&\times& [k^2 |\vec{q} - \vec{k}|^2 + 3 (\vec{k}\cdot{q} -k^2)^2][k^2 |\vec{q} - \vec{k}|^2 +  (\vec{k}\cdot{q} -k^2)^2].
\end{eqnarray}
Using the techniques already employed in section \ref{sec4} we can get
\begin{equation}
 {\mathcal F}_{\Pi}^{(FP)}(q) = \frac{13504 \pi }{1155}\ln{\biggl(\frac{\tau_{2}}{\tau_{1}}\biggr)}.
 \end{equation}
These results show indeed that ${\mathcal C}_{\rho_{\mathrm{gw}}}$,  ${\mathcal C}_{p_{\mathrm{gw}}}$ and  ${\mathcal C}_{\Pi_{\mathrm{gw}}}$ are all of the same order, as discussed in section \ref{sec4}. Similar results 
hold in the Landau-Lifshitz parametrization.
\end{appendix}

\end{document}